\documentclass[twocolumn]{aastex63}
\usepackage{amsmath}

\usepackage{hhline}

\begin{document}

\title{Exploration of Aspherical Ejecta Properties in Type Ia Supernova: Progenitor
Dependence and Applications to Progenitor Classification}

\author[0000-0002-4972-3803]{Shing-Chi Leung}	

  			  
\affiliation{TAPIR, Walter Burke Institute for Theoretical Physics, 
Mailcode 350-17, Caltech, Pasadena, CA 91125, USA}
  			  
\author[0000-0001-8304-7709]{Roland Diehl}  			  
  			  
\affiliation{Max-Planck-Institut f\"ur extraterrestrische Physik, Giessenbachstr. 
             1, 85741 Garching, Germany}
             
\affiliation{Excellence Cluster Universe, Technische Universit\"at M\"unchen, 
             Boltzmannstr. 2, D-85748, Garching, Germany}
  			  
\author[0000-0001-9553-0685]{Ken'ichi Nomoto}

\affiliation{Kavli Institute for the Physics and Mathematics of the Universe (WPI),
   			  The University of Tokyo Institutes for Advanced Study, The 
  			  University of Tokyo, Kashiwa, Chiba 277-8583, Japan}

\author[0000-0002-0552-3535]{Thomas Siegert}
           
\affiliation{Center for Astrophysics and Space Sciences, University of California, 
         	 San Diego, 9500 Gilman Dr, La Jolla, CA 92093, USA}

   \date{Received, accepted}
   
\shortauthors{Leung, Diehl, Nomoto and Siegert}
\shorttitle{Aspherical SN Ia: Progenitor Dependence and Applications}
   
\correspondingauthor{Shing-Chi Leung}
\email{scleung@caltech.edu}
   
\received{23 September 2019}  
\revised{2 April 2020}
\accepted{10 November 2020}   
\published{12 March 2021}
\submitjournal{Astrophysical Journal}

\date{\today}   
   
\begin{abstract}

Several explosions of Type Ia supernovae (SNe Ia) have been found to exhibit deviations
from spherical symmetry upon closer inspection. Examples are the gamma-ray lines from SN 2014J as
measured by INTEGRAL/SPI, and morphology information from radioactive isotopes in 
older remnants such as Tycho. 
A systematic study on the effects of parameters such as ignition geometry 
and burning morphology in SNe Ia is still missing. We use a 2D hydrodynamics code with post-processing
nucleosynthesis and simulate the double detonations in a sub-Chandrasekhar mass carbon-oxygen white dwarf
starting from the nuclear runaway in the accumulated He envelope towards disruption of the white dwarf.
We explore potential variety through four triggering scenarios that sample main asymmetry drivers.
We further investigate their global effects on the aspherical structure of the ejecta based on individual elements. 
We apply the results to the well observed SN 2014J and other recently
observed SN remnants in order to illustrate how these new observational data together with other observed 
quantities help to constrain the explosion and the progenitors of SNe Ia.

\end{abstract}


%

\section{Introduction}

\subsection{Observational Constraints on Type Ia Supernova Progenitors}

Type Ia supernovae (SNe Ia) have demonstrated a wide diversity \citep[e.g.,][]{Taubenberger2017}.
To trace the progenitor and the explosion mechanisms, 
supernova (SN) light curves and spectra are indispensable.
\citep[See e.g.,][and the Appendix for the summary of the progenitors and the explosion mechanism.]{Hillebrandt2000, Maoz2014, Nomoto2018SSR}
Explosion features can be extracted from the light curve shape \citep[e.g.,][]{Blondin2018},
spectral lines such as Ca II and Ni II \citep[e.g.,][]{Wilk2018}
and the polarization \citep[e.g.,][]{Bulla2016}.
The element abundance ratio can be the discriminant of current explosion models \citep[e.g.,][]{Seitenzahl2009},
by indicating the isotopic abundance ratios (e.g. $^{57}$Ni$/^{56}$Ni,
$^{55}$Fe/$^{56}$Ni, $^{44}$Ti/$^{56}$Fe) and element 
abundance ratios \citep[e.g., Mn/Fe and Ni/Fe; see][]{Mori2018}.
These techniques had been applied to a few well-observed
SNe Ia including SNe 2011fe, 2012cg, 2014J, 2015F and 
supernova remnants (SNR) 3C 397 (see recent works in for example
\cite{Yamaguchi2015,Leung2017NIC,Dave2017,Shen2018,Leung2018Chand,Zhou2020}). 
The progenitor mass and metallicity are constrained by this method,
which allows further identification of isotopic mass fractions
of radioactive Ni isotopes in SN 2012cg \citep{Graur2016} and SN 2014J \citep{Yang2018},
and stable Ni by the nebular IR spectra \cite[see][for the application to SN 2014J]{Dhawan2018}.

Radioactive isotopes with a longer half life, 
e.g., $^{44}$Ti with a half year of 59 years, can help to distinguish the 
different subclasses of SNe and their explosion mechanisms, 
by their effects on the decline rate of the late time light curve and the 
measuring the absolute abundances \cite[see e.g.,][]{Fry2015}).

Based on the tomography and morphology of SNR, 
the progenitor constraints are cast on SNR Tycho \citep{Wang2014,Lopez2015}.
The S and Fe spectral lines obtained from SNR
0519-69.0, 0509-67.5, and N103B are further examined in 
the shock-heated nebulae \citep{Seitenzahl2019}. The heated matter can emit
X-ray lines which provide direct constraints on the explosion 
mechanism, in particular the asphericity of SNe Ia.

Analysis of SNRs
on larger scales including the Small and Large Magellanic Cloud
can reveal the SN explosion history \citep{Maggi2016,Maggi2019}.
Study of particular elements e.g., Mn, can point out the 
major explosion mechanism \citep{McWilliam2018, delosReyes2020, Kobayashi2019}.

The morphology reveals the inherent explosion asymmetry.
For individual SNRs, 
the possible explosion progenitors can be inferred from their 
X-ray emission, which reveals the chemical abundances
\citep{Badenes2005,Badenes2006,Badenes2008}.
Differences in the interaction between the SN ejecta and the 
ambient medium can also point out the progenitor \citep{MartinezRodriguez2018}.
History of ejecta interaction with circumstellar matter (CSM) is 
contained in the late-time light curve as well
\footnote{We refer the interested
readers to the related presentations in the conference 
"Progenitors of Type Ia Supernovae" available 
in http://bps.ynao.cas.cn/xzzx/201908/t20190820\_510006.html}.

\subsection{Our Previous Studies of SNe Ia and Present Work on Asymmetries}

We have performed nucleosynthesis surveys of SN Ia explosions by the near-Chandrasekhar
mass white dwarf model using the turbulent deflagration model
with the deflagration-detonation transition in \cite{Leung2018Chand} (Paper I), 
the sub-Chandrasekhar mass white dwarf model using the double 
detonation model in \cite{Leung2018SubChand} (Paper II), and
the near-Chandrasekhar mass model for SNe Iax using the pure turbulent
deflagration model \cite{Leung2020Iax} (Paper III).
In these papers, we have conducted an extended parameter survey
or the SN Ia models aiming at understanding the implication of 
its observed diversity to progenitor model diversity, as well
as constraints on explosion physics. 

In Paper I, we considered models with the mass $M$ of 1.30 -- 1.38 $M_{\odot}$
and metallicity of $Z = 0 - 5 ~Z_{\odot}$. The super-solar metallicity is 
suggested to explain the observed SNe Ia, including the SN remnant 3C 397 
($Z \sim 5 Z_{\odot}$) and SN 2012cg ($Z \sim 3 - 5 Z_{\odot}$).
In Paper II, we consider WD models with $M = 0.9 - 1.3 M_{\odot}$,
$Z = 0 - 5 ~Z_{\odot}$, and the He envelope mass of $M_{\rm He} = 0.05 -- 0.2 M_{\odot}$.
We show that sub-Chandrasekhar mass models can also explain 
the isotopic ratio of nearby SNe Ia as the near-Chandrasekhar mass model does, 
but its Mn production cannot explain the nearby [Mn/Fe] trend taken 
from stars in the solar neighbourhood. Explosions of WD models with 
$M \sim 1.2 ~M_{\odot}$ 
can provide the key to distinguish the two explosion channels. 
In Paper III, we specifically model SNe Iax with the pure turbulent deflagration
modes of the WDs for masses of 1.30 -- 1.39 $M_{\odot}$ and $Z = Z_{\odot}$
\cite{Leung2020Iax}.

In the present paper IV, we study the asymmetry of the WD models
as suggested from SN 2014J, SNR Tycho, and other features mentioned above.
The ejecta geometry is primarily dependent on the explosion progenitor.
The near-Chandrasekhar mass model exploding by the turbulent deflagration
model with the deflagration-detonation transition tends to explode spherically
\cite[see, e.g.,][for recent three-dimensional realizations]{Roepke2007a}.
On the contrary, the sub-Chandrasekhar mass model
which explodes by the He-induced double detonations can generate 
large-scale asymmetry because of the off-center trigger of the explosion
\cite[see, e.g.,][for some recent three-dimensional realizations
showing aspherical structures]{Moll2013,GarciaSenz2018,Gronow2020}.

We thus focus on the sub-Chandrasekhar mass models, which 
tend to have a more aspherical structure than the near-Chandrasekhar mass models.
We note that in the literature there
is no extensive work examining how sub-Chandrasekhar mass models
exhibit a large-scale asphericity.
In this Paper IV, we study for the first time how the 
different detonation mechanisms of the sub-Chandrasekhar mass 
model can generate the ejecta deviated from the canonical spherical model
by multi-dimensional simulations.

Specifically, we will clarify how the detonation triggered in the 
He-envelope affects the $^{56}$Ni distribution and ejecta structure in
both position and velocity spaces,
We also try to understand the underlying principles for the observed
irregularities in SNRs.

We choose two-dimensional models
so that we may test a larger number of models than three-dimensional models to uncover the 
global trend of the parameter landscape. Also, quasi-spherical SN 2014J (see \S\ref{sec:SN2014J}) has
encouraged us to explore models with a certain level of symmetry (e.g., rotation
symmetry and reflection symmetry) assumed in two-dimensional model, 
instead of arbitrary models without explicit symmetry 
as in three-dimensional models.

Ideally, three-dimensional models are naturally
desired to match all features self-consistently. However, they are much more
computationally expensive. As an exploratory study, we aim at 
searching the key properties in the detonation setting for the
models to contain different features that might ultimately imprint in observational data.
We want to understand what kind of shock interaction is necessary
for generating the observed features, from which we may obtain
hints on the initial detonation pattern. This will guide future 
three-dimensional simulations in setting up accurate initial models
aiming for explaining SN 2014J or other supernovae.

\subsection{Paper Structure}

In Section \ref{sec:methods}, we describe our methodology
and the models to be presented in this article. Then, we briefly review how we compute the 
explosion models of the sub-Chandrasekhar mass WD as the progenitor
and we present the stellar parameters, explosion energetics, 
and the essential nucleosynthetic products.

In Section \ref{sec:representation}, we first present how 
the near-Chandrasekhar and sub-Chandrasekhar mass WDs differ
by their large-scale asymmetry. Then we examine in details
how the explosion ejecta and its chemical composition depend
on the viewing (ejecta) angle. We also predict the expected
morphology by extracting the representative elements.

In Section \ref{sec:SN2014J}, we present a detailed case 
study for SN 2014J to show how the asymmetry and other nucleosynthetic
yields can be used for constraining the explosion mechanism
and the progenitors of SNe Ia. We cast constraints on the progenitor
mass, initial explosion geometry, He envelope mass, and 
its metallicity. 

In Section \ref{sec:extension}, we further apply our results
on some recently observed SN remnants reported in the literature
to demonstrate how the geometry can provide us the hints.

In Section \ref{sec:discussion}, we discuss how this work is 
related to other hydrodynamics simulations in the literature.

\section{Methods and Models}
\label{sec:methods}

We use our two-dimensional hydrodynamics code developed for modeling 
the explosion models in this work. The code is based on high-order
shock capturing scheme and time-discretization scheme, coupled with 
sub-grid scale turbulence models, flame tracking schemes and nuclear reaction networks
of arbitrary sizes. We refer the interested reader to the instrumentation
paper which describes the prototype \citep{Leung2015a}. We have further 
extended the code to accommodate the code in different explosion scenarios.
Different extensions are described in details for

\noindent (1) SNe Ia in
\cite{Leung2017NIC,Leung2018Chand,Leung2018SubChand}, 

\noindent (2) electron capture SNe in \cite{Leung2018ECSN,Leung2017ECSN_Proceeding,Leung2019,Zha2019b},
and 

\noindent (3) dark matter admixed compact objects in \cite{Leung2015b,Leung2019DMAIC,Zha2019a}.

\subsection{Input Physics}

We follow \cite{Leung2018SubChand} for SN Ia using the sub-Chandrasekhar mass WD models.
We solve the two-dimensional Euler equations in cylindrical coordinates.
We further assume reflection symmetry of the $x$-$y$ plane so that we model
only one quadrant of the star. 
We use a realistic Helmholtz EOS \citep{Timmes1999a} for describing the 
matter with free electrons with arbitrary relativistic level and degeneracy,
nuclei as a classical ideal gas, photons with Planck distribution and
electron-positron pair effects. 
Different from the near-Chandrasekhar mass WD models, 
 we include:

\noindent (1) the notation of He-detonation in the simulation, 

\noindent (2) its energy generation prescription and timescale, and 

\noindent (3) its propagation velocity. 

We use the same solver for matter in the nuclear statistical equilibrium (NSE)
as it is independent of the original composition of the matter, and only depends on
the final density, temperature and electron fraction. Level-set methods are
used for tracing the contour of the He- and C-detonation fronts. 
In Paper II, we have further performed a set of tests to justify the 
functionality of the code in the appendix. They include that

\noindent (1) the C-detonation trigger is independent of the symmetry boundary we used\footnote{We explored whether modeling the 
two-bubble structure using a hemisphere and a quadrant gives
rise to different results. We showed that indeed the detonation waves collide identically as if they are laminar wave
at the boundary where reflective 
boundary is imposed. In fact, this scenario provides a more stringent 
test to how robust the second detonation can be triggered because
there is no geometric convergence taking place near the reflective boundary. 
As a result, the required He envelope mass predicted by this assumption
is the upper limit. This value can be drastically reduced as geometric
effects become stronger.},

\noindent (2) the explosion energetics are insensitive to the resolution used,

\noindent (3) the shock convergence is less sensitive to the resolution, and

\noindent (4) the threshold of the detonation trigger is independent on the resolution.

We have further shown that in Paper II that our two-dimensional models give agreeing
results with contemporary one-, two- and three-dimensional models found in the 
literature.

\subsection{Post-Process Nucleosynthesis}

In order to keep track of the detailed nucleosynthesis for constructing the 
isotope distribution, we use the tracer particle scheme 
\citep{Travaglio2004,Seitenzahl2009,Townsley2016}. 
It makes use of the massless Lagrangian tracers. They follow the fluid motion and 
record their own thermodynamical trajectories. The particles are "massless" 
that they do not change the fluid motion. After the hydrodynamical simulations,
the tracers are post-processed with a large nuclear reaction network \citep[a 495-isotope
network containing isotopes from $^1$H to $^{91}$Tc][]{Timmes2000a}.
After post-processing, the spatial distribution of specific isotopes
such as $^{16}$O, $^{28}$Si and $^{56}$Ni are extracted for this data analysis.

\subsection{Models}

In this work, based on the formalism of our previous works, we examine further
SN Ia models in the range between 0.95 -- 1.0 $M_{\odot}$ and we extend
systematically to different initial He-detonation structures. 
We evolve WDs from the onset of He-detonation until no significant
exothermic reactions take place.
This can be because the WD is disrupted completely by both types of 
detonation, or the He-detonation fails to trigger the second detonation
and quenches.
In Table \ref{table:models} we list initial densities 
and temperature used in our models. 
As indicated by observations data, we consider
the He-envelope mass $M_{{\rm He}}$ from 0.05 -- 0.10 $M_{\odot}$.

In the table, we also list other related parameters
including the central density $\rho_c$ and the interface
density $\rho_{{\rm He}}$. Among 
all our considered models, the densities range from $\sim 10^{7}$ to $10^{8}$ g cm$^{-3}$
while the He-interface is from $\sim 10^{5}$ to $10^{6}$ g cm$^{-3}$. 
Notice that this allows a major part of the star to carry out  
complete burning from CO matter to ashes in NSE
at a density $> 5 \times 10^7$ g cm$^{-3}$. 

In general, three types of events can
result. "N/A" means that no C-detonation occurs: this happens
when the He-detonation is too weak (without the possible 
geometric convergence) to heat the CO matter to the 
sufficient temperature, or collision to create the required penetration. 
Results "cen" and "off" 
stand for centered and off-centered detonation, respectively. 
We also list the yielded $^{56}$Ni and $^{57}$Ni 
masses, obtained at the end of simulations, where most
exothermic reactions have ceased.

\begin{figure}
\centering
\includegraphics*[width=8cm,height=5.7cm]{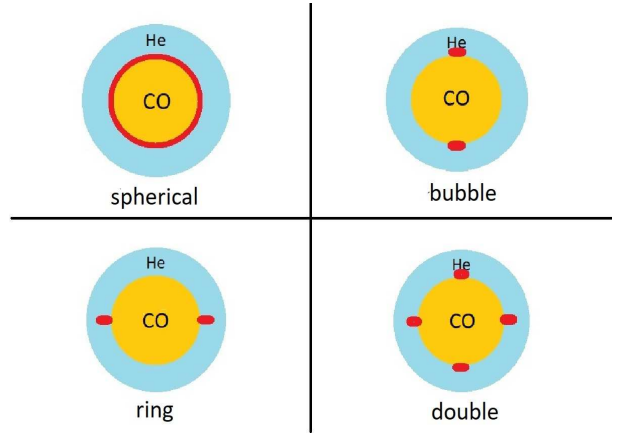}
\caption{Graphical illustrations of the initial He-detonation
configuration used in this article. The figure depicts
the cross-sections of initial CO WDs with He-envelopes
and the initial He-detonation put in by hand. The "S"- (spherical),
"B"- (bubble), "R"- (ring) and "D"- (double) Types of 
geometry are presented. The orange and light blue regions 
stand for the CO-rich and He-rich regions . The red region stands 
for the zone which is assumed to be burnt already 
at the beginning of the simulations.}
\label{fig:HeDet}
\end{figure}

In Figure \ref{fig:HeDet} we depict the four scenarios
used in this work. They include a bubble ("B"-Type), 
a ring ("R"-Type), a bubble and a ring ("D"-Type) and 
a sphere ("S"-Type). This spans the possible initial
He-runaway from the lowest symmetry to the highest symmetry. 
In the figure the cross-sections of the WD
progenitor are drawn. 
The "B"-Type
detonation corresponds to two bubbles, one at the "north"-pole.
and one at the "south"-pole.

In order to realize the one-bubble event, 
simulations modeling the hemisphere explicitly is necessary. 
We remind that the C-detonation is triggered is very similar
to the "R"-Type since, in this configuration, the C-detonation 
always starts after the shock converges at the opposite "pole"
from where the detonation is initialized.
Therefore, we may refer to "R"-Type series to trace how
the detonation takes place. We note that, computationally, 
the detonation trigger is identical as indicated by Appendix B in Paper II. 

\begin{table*}
\begin{center}
\caption{Initial conditions, explosion energetics and 
the global nucleosynthetic results of the sub-Chandrasekhar mass
SN Ia models presented in this work. 
Models in bold font text are those with a comparable $^{56}$Ni mass
with SN 2014J.
$M$, $M_{{\rm He}}$
and $M_{{\rm CO}}$ are the masses of the initial WD, 
He envelope, and CO core in units of solar mass. 
The column "flame" stands for the initial He-runaway geometry.
$R$ and $R_{{\rm He}}$ are the radii of the initial WD
and the interface from the CO core to the He envelope in units
of km. $\rho_c$ and $\rho_{{\rm He}}$ are the densities in the 
core and in the CO-He interface. The columns "runaway?" and "type"
represent whether the model develops the C-detonation and 
how the C-detonation is triggered (see also in the main text for 
the definition). "cen" and "off" stand for centered and off-centered
C-detonation, while N/A means that no second detonation occurred. 
$M({{\rm ^{56}Ni}})$ and $M({{\rm ^{57}Ni}})$ are 
the final $^{56}$Ni and $^{57}$Ni computed by post-processing
nucleosynthesis in units of $M_{\odot}$.}
\label{table:models}
\begin{tabular}{|c|c|c|c|c|c|c|c|c|c|c|c|c|}
\hline

Model & $M$ & $M_{{\rm He}}$ & $M_{{\rm CO}}$ & flame & $R$ & $R_{{\rm He}}$ & $\rho_c$ & $\rho_{{\rm He}}$ & $2^{{\rm nd}}$ runaway? & type? & $M({{\rm ^{56}Ni}})$ & $M({{\rm ^{57}Ni}})$ \\ \hline

095-050-B & 0.95 & 0.05 & 0.90 & B & 6710 & 4760 & 2.23 & 0.06 & No  & N/A & N/A  & N/A \\
095-050-R & 0.95 & 0.05 & 0.90 & R & 6710 & 4760 & 2.23 & 0.06 & Yes & off   & 0.11 & $3.04 \times 10^{-3}$ \\
095-050-D & 0.95 & 0.05 & 0.90 & D & 6710 & 4760 & 2.23 & 0.06 & No  & N/A & N/A  & N/A \\
095-050-S & 0.95 & 0.05 & 0.90 & S & 6710 & 4760 & 2.23 & 0.06 & Yes & cen   & 0.45 & $1.14 \times 10^{-2}$ \\ \hline
095-100-B & 0.95 & 0.10 & 0.85 & B & 6710 & 4330 & 2.23 & 0.12 & No  & N/A & N/A  & N/A \\ 
095-100-R & 0.95 & 0.10 & 0.85 & R & 6710 & 4330 & 2.23 & 0.12 & Yes & off   & 0.31 & $8.65 \times 10^{-3}$ \\
095-100-D & 0.95 & 0.10 & 0.85 & D & 6710 & 4330 & 2.23 & 0.12 & Yes & off   & 0.29 & $8.63 \times 10^{-3}$ \\
\textbf{095-100-S} & 0.95 & 0.10 & 0.85 & S & 6710 & 4330 & 2.23 & 0.12 & Yes & cen   & 0.48 & $1.28 \times 10^{-2}$ \\ \hline
100-050-B & 1.00 & 0.05 & 0.95 & B & 6180 & 4350 & 3.21 & 0.09 & No  & N/A & N/A  & N/A \\
100-050-R & 1.00 & 0.05 & 0.95 & R & 6180 & 4350 & 3.21 & 0.09 & Yes & off   & 0.31 & $8.16 \times 10^{-3}$ \\ 
100-050-D & 1.00 & 0.05 & 0.95 & D & 6180 & 4350 & 3.21 & 0.09 & Yes & off   & 0.08 & $2.31 \times 10^{-3}$ \\ 
\textbf{100-050-S} & 1.00 & 0.05 & 0.95 & S & 6180 & 4350 & 3.21 & 0.09 & Yes & cen   & 0.60 & $1.60 \times 10^{-2}$ \\ \hline
100-100-B & 1.00 & 0.10 & 0.90 & B & 6180 & 3980 & 3.21 & 0.16 & Yes & off   & 0.35 & $1.14 \times 10^{-2}$ \\
\textbf{100-100-R} & 1.00 & 0.10 & 0.90 & R & 6180 & 3980 & 3.21 & 0.16 & Yes & off   & 0.46 & $1.30 \times 10^{-2}$ \\
\textbf{100-100-D} & 1.00 & 0.10 & 0.90 & D & 6180 & 3980 & 3.21 & 0.16 & Yes & off   & 0.44 & $1.26 \times 10^{-2}$ \\
\textbf{100-100-S} & 1.00 & 0.10 & 0.90 & S & 6180 & 3980 & 3.21 & 0.16 & Yes & cen   & 0.62 & $1.74 \times 10^{-2}$ \\ \hline
105-050-B & 1.05 & 0.05 & 1.00 & B & 5300 & 4110 & 4.33 & 0.10 & No  & N/A & N/A  & N/A \\
\textbf{105-050-R} & 1.05 & 0.05 & 1.00 & R & 5300 & 4110 & 4.33 & 0.10 & Yes & off   & 0.48 & $1.24 \times 10^{-2}$ \\ 
\textbf{105-050-D} & 1.05 & 0.05 & 1.00 & D & 5300 & 4110 & 4.33 & 0.10 & Yes & off   & 0.48 & $1.37 \times 10^{-2}$ \\ 
105-050-S & 1.05 & 0.05 & 1.00 & S & 5300 & 4110 & 4.33 & 0.10 & Yes & cen   & 0.76 & $1.63 \times 10^{-2}$ \\ \hline
\textbf{105-100-B} & 1.05 & 0.10 & 0.95 & B & 5300 & 3730 & 4.33 & 0.19 & Yes & off   & 0.49 & $1.65 \times 10^{-2}$ \\
\textbf{105-100-R} & 1.05 & 0.10 & 0.95 & R & 5300 & 3730 & 4.33 & 0.19 & Yes & off   & 0.59 & $1.78 \times 10^{-2}$ \\
\textbf{105-100-D} & 1.05 & 0.10 & 0.95 & D & 5300 & 3730 & 4.33 & 0.19 & Yes & off   & 0.59 & $1.80 \times 10^{-2}$ \\
\textbf{105-100-S} & 1.05 & 0.10 & 0.95 & S & 5300 & 3730 & 4.33 & 0.19 & Yes & cen   & 0.70 & $2 \times 10^{-2}$ \\ \hline
110-050-B & 1.10 & 0.05 & 1.05 & B & 4930 & 3800 & 6.17 & 0.13 & No  & N/A & N/A  & N/A \\
\textbf{110-050-R} & 1.10 & 0.05 & 1.05 & R & 4930 & 3800 & 6.17 & 0.13 & Yes & off   & 0.68 & $1.90 \times 10^{-2}$ \\ 
\textbf{110-050-D} & 1.10 & 0.05 & 1.05 & D & 4930 & 3800 & 6.17 & 0.13 & Yes & off   & 0.60 & $1.72 \times 10^{-2}$ \\ 
110-050-S & 1.10 & 0.05 & 1.05 & S & 4930 & 3460 & 6.17 & 0.13 & Yes & cen   & 0.82 & $1.90 \times 10^{-2}$ \\ \hline
\textbf{110-100-B} & 1.10 & 0.10 & 1.00 & B & 4930 & 3460 & 6.17 & 0.24 & Yes & off   & 0.61 & $2.10 \times 10^{-2}$ \\
110-100-R & 1.10 & 0.10 & 1.00 & R & 4930 & 3460 & 6.17 & 0.24 & Yes & off   & 0.75 & $2.37 \times 10^{-2}$ \\
\textbf{110-100-D} & 1.10 & 0.10 & 1.00 & D & 4930 & 3460 & 6.17 & 0.24 & Yes & off   & 0.70 & $2.21 \times 10^{-2}$ \\
110-100-S & 1.10 & 0.10 & 1.00 & S & 4930 & 3460 & 6.17 & 0.24 & Yes & cen   & 0.81 & $2.43 \times 10^{-2}$ \\ 
\hhline{|=|=|=|=|=|=|=|=|=|=|=|=|=|} 
Observations & & & & & & & & & & & & \\ \hline
lower limit & 0.7 & 0.03 & & & & & & & & & 0.4 & $2.32 \times 10^{-2}$ \\ 
upper limit & 3.1 & 0.09 & & & & & & & & & 0.7 & $5.53 \times 10^{-2}$ \\ \hline

\end{tabular}
\end{center}
\end{table*}

\section{Representation of Asphericity of SNe Ia}
\label{sec:representation}

In this section we examine how the asphericity of a SN Ia 
can be embodied by their observables. We first compare the
typical ejecta structure from our two-dimensional simulations
for the near-Chandrasekhar and sub-Chandrasekhar models 
to see how the progenitor mass affects the asphericity.
Then we focus on the sub-Chandrasekhar mass model to 
show how the initial detonation affects the ejecta
distribution, velocity and remnant morphology.

\subsection{Near-Chandrasekhar Mass WD versus sub-Chandrasekhar Mass WD}

\begin{figure}
\centering
\includegraphics*[width=8cm]{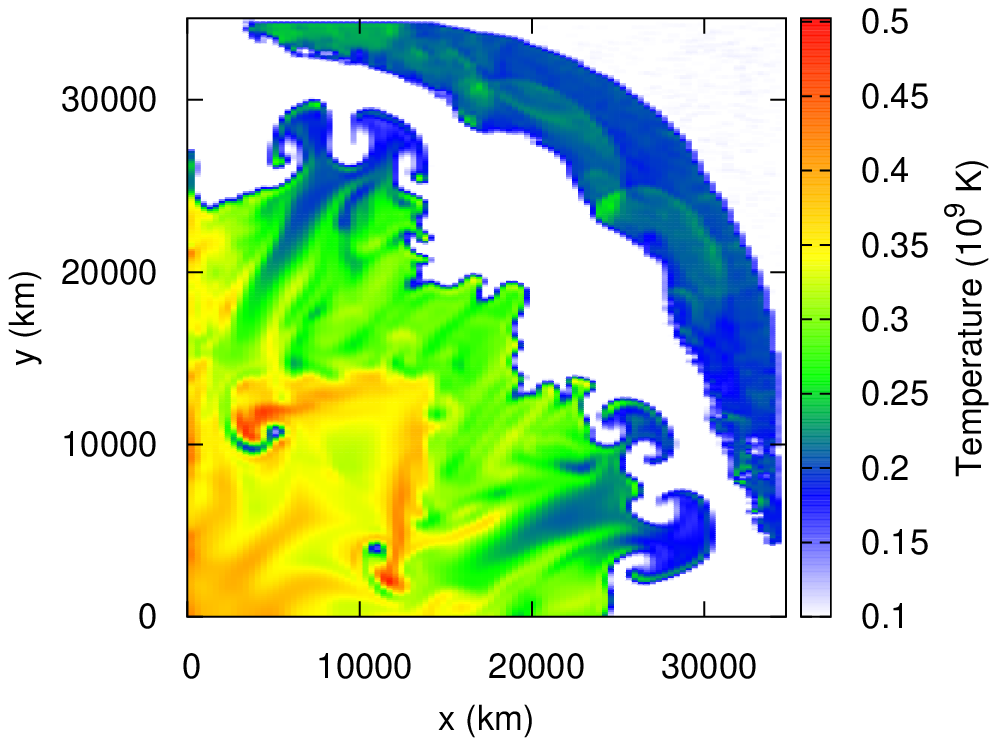}
\includegraphics*[width=8cm]{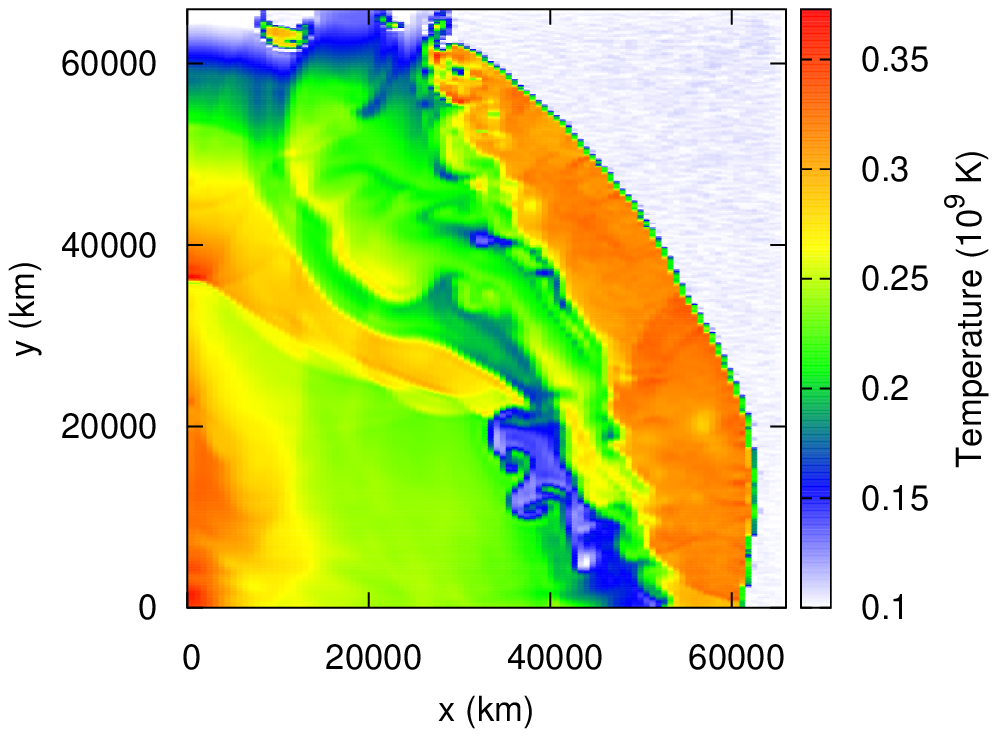}
\caption{(top panel) The colour temperature profile of our benchmark
SN Ia models from the explosion of a near-Chandrasekhar mass WD
at $\sim 4$ s after the nuclear runaway has started (see Paper II
for more details). (bottom panel)
Similar to the top panel, but from a sub-Chandrasekhar Mass WD (this work).}
\label{fig:ChandvssubChand}
\end{figure}

In this work we focus on the asymmetry of sub-Chandrasekhar mass
WD. We do not focus on the near-Chandrasekhar mass WD because we observe
that typical SN Ia models from the near-Chandrasekhar mass WD tends to
have a more spherical structure. Here we present a comparison 
to outline the large-scale asymmetry in both models. 
For details of these models we refer interested readers
to Papers I, II, and III for more detailed
description and implementation. 

The two models are chosen from our benchmark models. Both 
models are chosen to represent the "normal" SNe Ia by its
$^{56}$Ni production $\sim 0.6 ~M_{\odot}$. 
Furthermore the near-Chandrasekhar mass model also
produces the necessary amount of $^{55}$Mn sufficient for 
reproducing the trends of stellar abundance
near the solar neighbourhood.

In the top panel of Figure \ref{fig:ChandvssubChand}
we show the temperature in a section 
through the star of the near-Chandrasekhar 
mass model at $\sim 4$ s after the explosion.
The model assumes turbulent deflagration with 
deflagration-detonation transition. The initial nuclear
runaway assumes a centered deflagration wave with angular
perturbations.
The near-Chandrasekhar mass WD has a mass 1.37 $M_{\odot}$
and a central density $3 \times 10^9$ g cm$^{-3}$.
A centered-flame $c3$ (with three-finger structure) 
is used as the initial nuclear runaway. The finger structure
is used to enhance hydrodynamical instabilities for
the asphericity. 
The sub-Chandrasekhar mass model has a total mass 1.1 $M_{\odot}$
with a helium envelope mass $0.1 ~M_{\odot}$ and
a central density $\sim 6 \times 10^7$ g cm$^{-3}$. 
A one-bubble configuration is placed along the rotation axis.

For our model, at $t \sim 4 - 5$ s the global distribution
of ejecta structure begins to be frozen. Secondary features
including Rayleigh-Taylor instabilities ("mushrooms")
still continue to grow. 
But most features are smoothed by the expansion,
leaving a surface close to spherical. 
We can observe that by the external detonation transition,
the detonation wave always wraps around the aspherical ash produced
during deflagration. Surface asphericity in temperature 
appears to be significant in the Chandrasekhar mass model 
at a radius $\sim 30000$ km. 
However, the matter at the surface is mostly C and O. 
The area of interest, where iron-peak elements are synthesized, does not 
show much asphericity. They are produced primarily in the
core, which means it is unlikely for large-scale asymmetry
features from $^{56}$Ni and $^{56}$Co can be generated
and be exposed. In this work, the core corresponds to the CO-rich matter under the He-envelope.

In the bottom panel of Figure \ref{fig:ChandvssubChand}, we show the temperature profile 
of a sub-Chandrasekhar mass CO white dwarf with a 
He envelope for comparison.
The sub-Chandrasekhar mass model 
show a more explicit large-scale asphericity. 
The detonation is triggered near the "equator". 
As a result, the He-detonation is stronger near the 
"equator", as shown by the hot spot in 60000 km
closer to the symmetry axis. Since the C-detonation
propagates from the place near "equator" to the 
center and then outward. The inner ejecta from
C-detonation has a preferred direction along the rotation
axis. The high velocity flow along this direction will
be responsible for the later large-scale asphericity.

\subsection{Detonation Geometry Induced Asphericity}

\begin{figure*}
\centering
\includegraphics*[width=18cm,height=5.7cm]{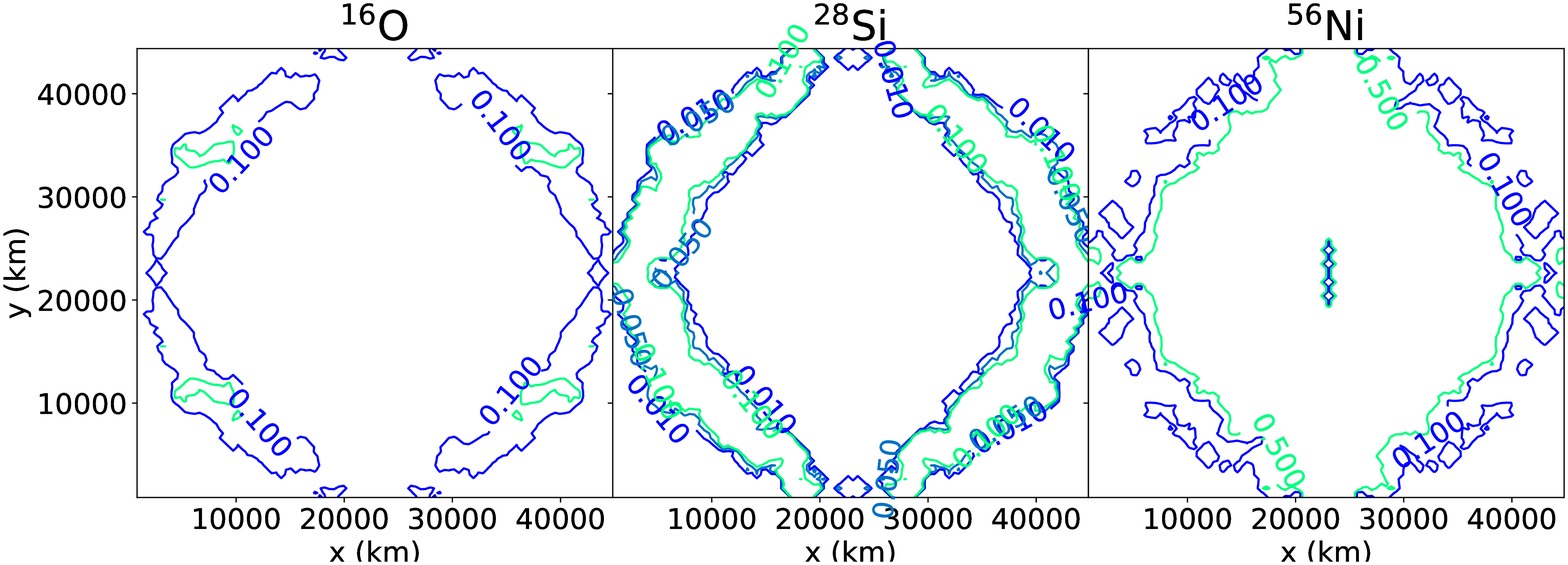}
\includegraphics*[width=18cm,height=5.7cm]{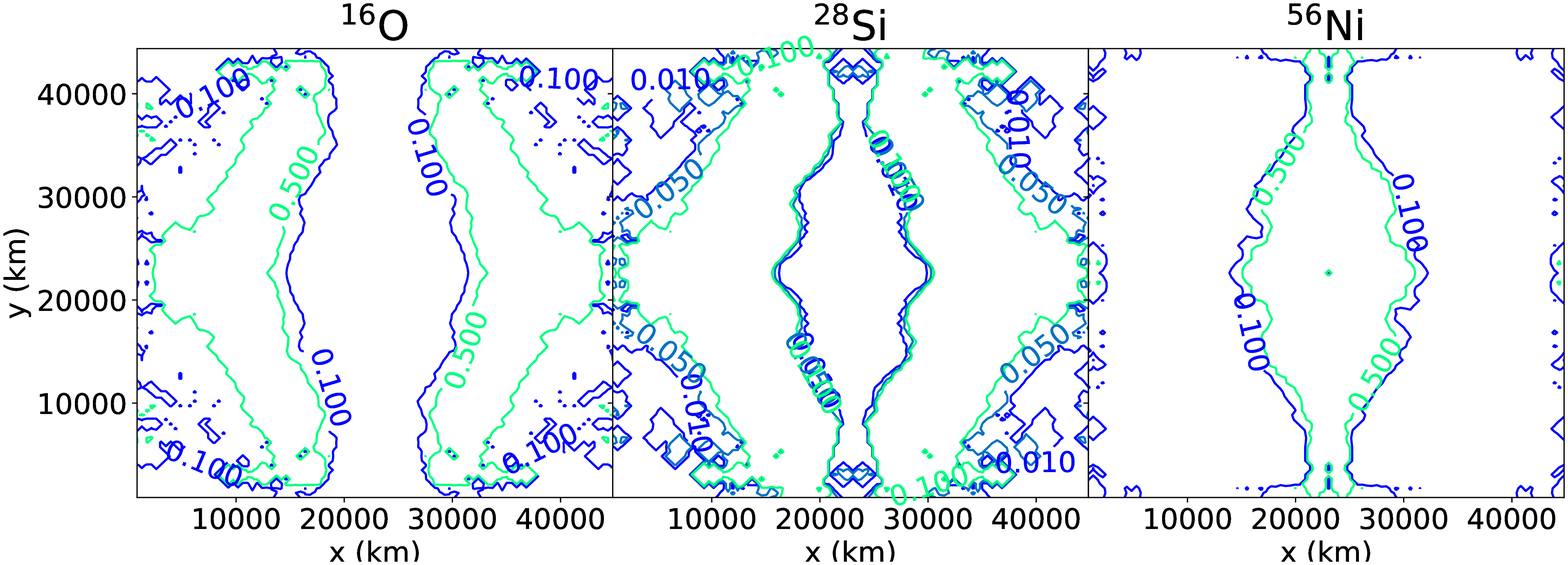}
\includegraphics*[width=18cm,height=5.7cm]{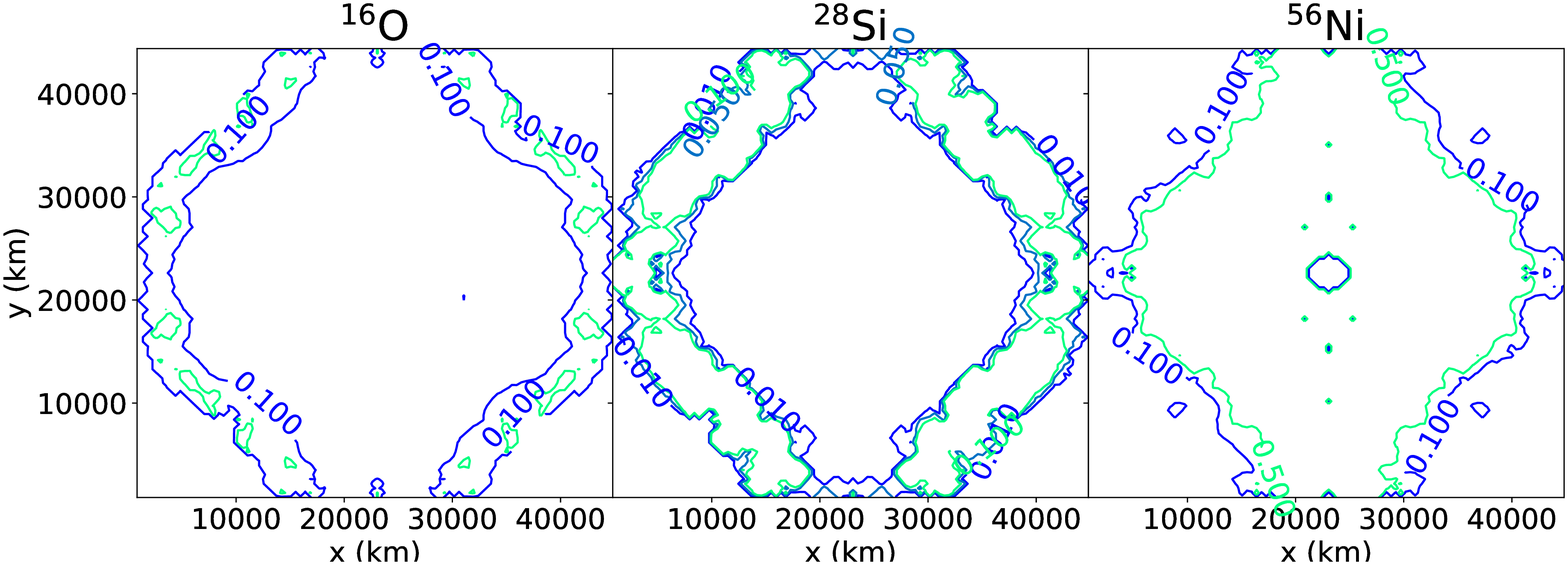}
\includegraphics*[width=18cm,height=5.7cm]{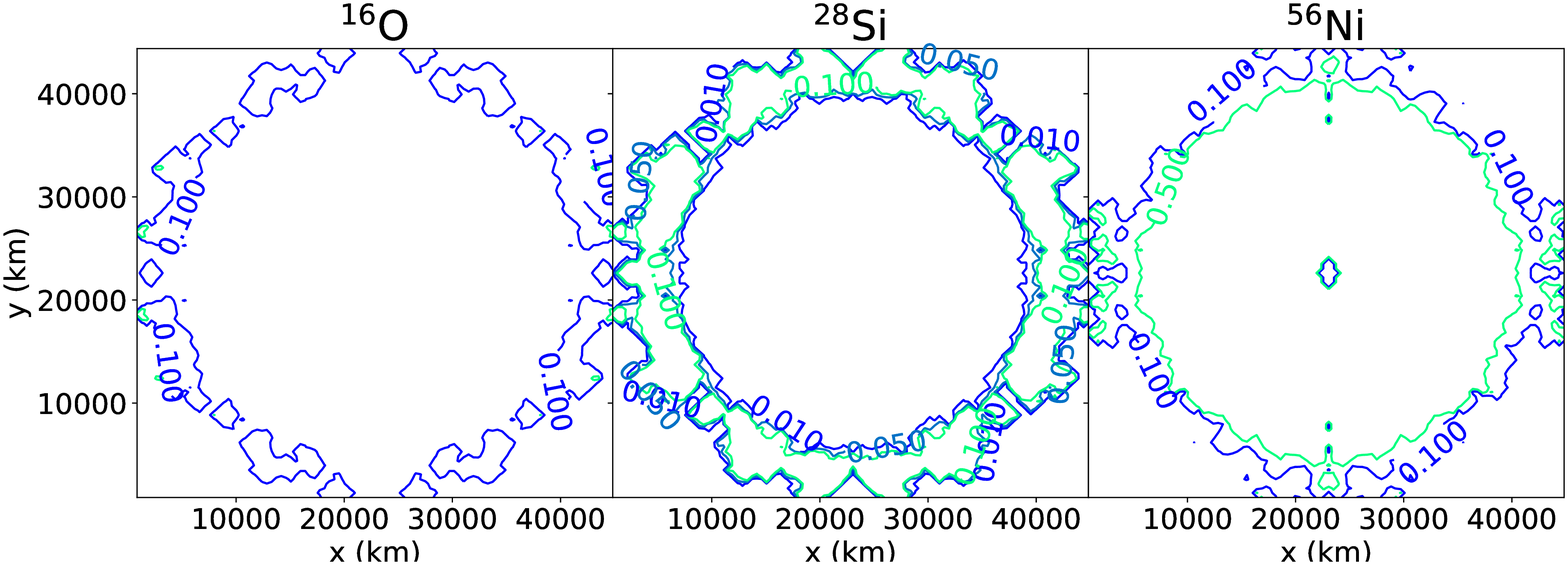}
\caption{Mass fraction distributions of $^{16}$O, $^{28}$Si and $^{56}$Ni for the 
explosion models 100-100-R (top panel), 100-100-B (second top panel), 
100-100-D (third) and 100-100-S (bottom panel).
The numbers stand for the corresponding contours of the 
mass fraction for the particular isotope.}
\label{fig:NiDist}
\end{figure*}

Having shown that the sub-Chandrasekhar mass WD is capable of 
generating large-scale asymmetry, we further examine 
how the distribution of $^{56}$Ni deviates from spherical symmetry. 
In Figure \ref{fig:NiDist} we show the final
distributions of the tracer particles which are the 
mass fraction contours of $^{16}$O, $^{28}$Si and $^{56}$Ni. As examples, we consider
models 100-100-R, 100-100-B and 100-100-S.

For Model 100-100-S, we can see that all the Ni-rich matter
is covered by the thick envelope. 
We choose this series because
all these models produce a comparable $^{56}$Ni mass 
as a normal SN Ia.

The "S"-Type model does not show any significant
amount of $^{56}$Ni near the surface. The distribution confirms that 
the initial spherical symmetry makes the detonation wave propagate only spherically.
Except for small scale perturbation coming from discretization
effects from cylindrical coordinate to spherical coordinate, 
the large-scale distribution is to a good approximation spherical. 
The detonation waves do not collide with each other
except at the stellar core when the detonation waves converge. 
There is no detonation wave interaction by shock collision
or shock convergence in the He envelope. 
The amount of $^{56}$Ni
production becomes lower in the envelope. Hence, this 
model is less likely to explain the observed early
$^{56}$Ni decay line. 

The "R"- and "B"-Type models show more 
near surface $^{56}$Ni. The $^{56}$Ni is ejected more 
along the polar direction for the "B"-Type detonation,
meanwhile the "R"-Type detonation ejects matter more
spherical,
but in a more elongated manner compared with "S"-Type model.
The "D"-Type model also behaves similarly to the "R"-Type
except that the $^{56}$Ni around the "equator" is lower in the 
abundance. 

We remind that it is the asynchronous burning of helium,
coupled with the geometric convergence creates the observed
asphericity. The asynchronous burning allows He with the same
initial density, to be burnt at different time. When the 
detonation time is delayed, the density of the He-matter
decreases, thus making it less likely to generate $^{56}$Ni
in the first place. On the contrary, the geometric convergence
creates the hot spot for triggering $^{56}$Ni synthesis robustly.
However, the effect of geometric convergence is more localized. 
In a similar way, the detonation wave collision can create 
the high temperature and high density
zones for synthesizing $^{56}$Ni.

\subsection{Velocity Distribution of Ejecta}

\begin{figure*}
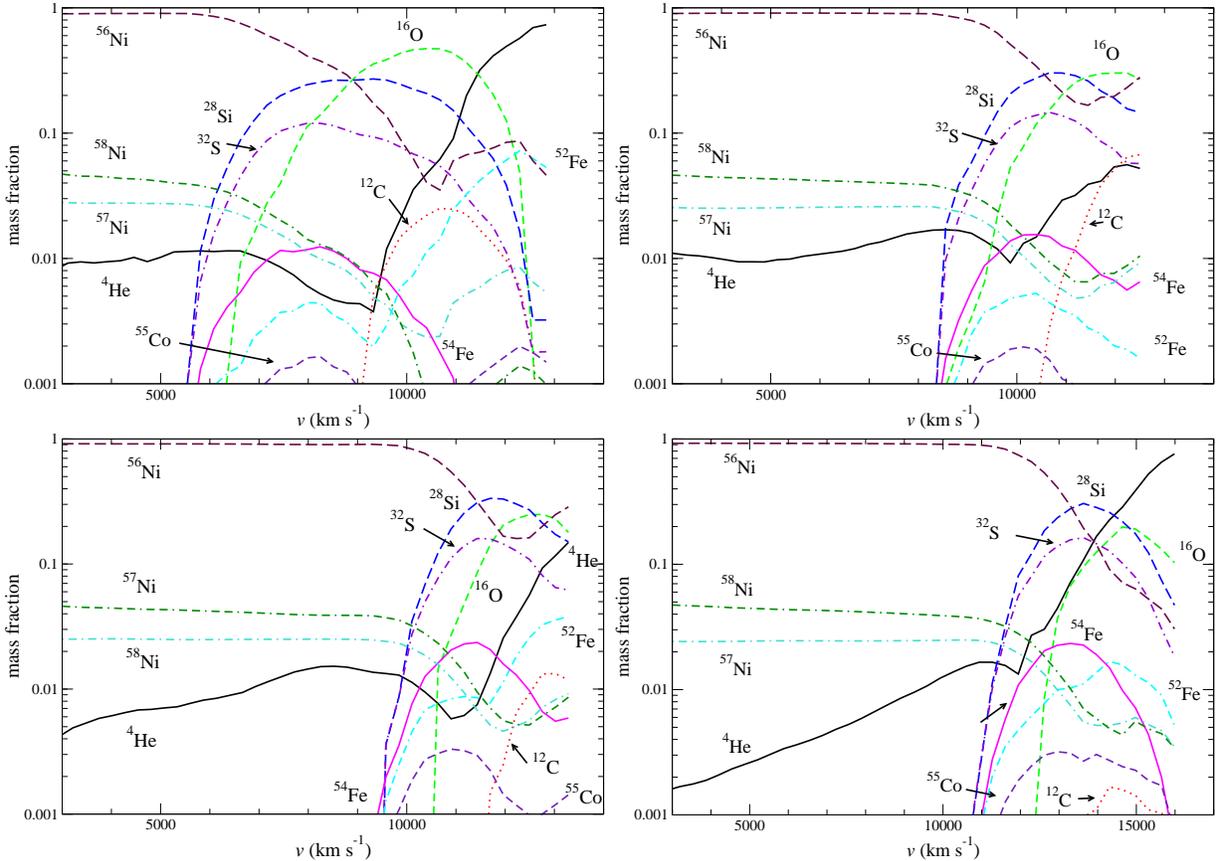

\centering
\includegraphics*[width=8cm,height=5.7cm]{fig4a.eps}
\includegraphics*[width=8cm,height=5.7cm]{fig4b.eps}
\includegraphics*[width=8cm,height=5.7cm]{fig4c.eps}
\includegraphics*[width=8cm,height=5.7cm]{fig4d.eps}

\caption{Velocity distributions of the ejecta for Models
100-100-B (top left), 100-100-D (top right), 100-100-R (lower left)
and 100-100-S (lower right) respectively.}

\label{fig:vel}
\end{figure*}

One of the features in SN2014J is the early time $^{56}$Ni signal
and time-dependent velocity for late-time $^{56}$Co (see \S\ref{sec:SN2014J}).
It is therefore interesting to further study how the initial 
detonation configuration can give rise to the diversity of 
the isotope velocity distributions. In Figure \ref{fig:vel}
we plot the ejecta distribution in the velocity space to 
analyze how the detonation affects the final ejecta distribution. 
The distribution is angular-averaged. More samples of SNe Ia show
similar early pre-maximum bumps (see, e.g., \cite{Jiang2017,Jiang2018}).
The possibilities of these SNe Ia forming a sub-class showing
that SN Ia with observable He-burning features can have a 
common evolutionary path. Furthermore, such early gamma-ray flux can be another
important sign for future telescopes to capture the early optical
evolution of these SNe \citep{Wang2019}. 

For Model 100-100-B, the one bubble configuration allows ejecta to be
concentrated with $^{56}$Ni-rich material (see Figure \ref{fig:NiDist})
near the "equator". As a result, there is a multi-layered distribution
of $^{56}$Ni. Below 6000 km s$^{-1}$, the ejecta are filled with 
$^{56-58}$Ni. From 6000 to 8000 km s$^{-1}$, $^{28}$Si and $^{32}$S
are the major isotopes. 8000 -- 11000 km s$^{-1}$ unburnt oxygen is the 
major element and outside He is in the main element in the ejecta.
We remark that the $^{56}$Ni distribution is not monotonically
decreasing, as compared to the classical spherical model. It first drops 
around 10000 km s$^{-1}$, showing that the detonation reaches 
the low-density region for C-burning. After that the mass 
fraction of $^{56}$Ni rises again and quickly drops in 
its abundance. This shows that the shock collision in the He-envelope
allows formation of $^{56}$Ni directly. But the shock strength 
is not strong enough to channel the outburst of $^{56}$Ni,
as seen by the covering layer of $^{4}$He. 

For Model 100-100-D, the strong collision away from the axis 
allows an outburst of $^{56}$Ni during $^{4}$He-burning at
early time. This is also reflected in the ejecta distribution. 
Ejecta with a velocity below 10000 km s$^{-1}$ are again filled 
with Ni isotopes. $^{28}$Si and $^{32}$S are the major
isotopes in the velocity range 10000 -- 11000 km s$^{-1}$.
Products of incomplete C-burning, such as unburnt $^{16}$O, 
can be found most abundant up to 12000 km s$^{-1}$. Outside
that $^{56}$Ni and $^{16}$O are the major isotopes. 
From this it can be seen that multiple ignitions allow
$^{4}$He to be burnt quickly along the $\alpha$-chain.
Such outermost $^{56}$Ni can be readily ejected and be
seen through its decay. 

For Model 100-100-R, the geometric convergence takes place at the
pole which is strong enough to create similar pinching to the 
He-envelope. The ejecta are covered with IMEs
from 10000 -- 11000 km s$^{-1}$ and $^{16}$O from 11000 to 12000 km s$^{-1}$. 
The outermost layer is mixed with $^{56}$Ni and $^{4}$He with 
hints of $^{28}$Si and $^{32}$S. 
Similar to Model 100-100-D, a thin layer of 
IMEs between 10000 -- 11000 km s$^{-1}$
and then incomplete C-burning products between 11000 -- 12000 km s$^{-1}$. 
The outermost ejecta are a mixture of $^{56}$Ni, $^{4}$He, 
$^{28}$Si, $^{32}$S and $^{16}$O. 

For Model 100-100-S, the spherical detonation allows a stratified 
structure in the ejecta. Ejecta with a velocity below 
12000 km s$^{-1}$ are dominated by $^{56}$Ni, $^{57}$Ni
and $^{58}$Ni. IMEs including $^{28}$Si and
$^{32}$S are mostly found between 12000 -- 14000 km s$^{-1}$.
Ejecta with a velocity $> 14000$ km s$^{-1}$ are occupied by $^{4}$He.

By comparing these four models, it demonstrates the possibility of 
mixing $^{56}$Ni at high velocity, namely at the outermost ejecta
for the "D" and "B"-Types.
The oblique shock and the geometric convergence of detonation inside
the He-envelope can provide the necessary thrust for channeling the 
$^{56}$Ni produced in the He-detonation to the outermost part of 
ejecta. When the ejecta quickly expands, the gamma-rays produced by 
the decay of $^{56}$Ni into $^{56}$Co may be directly seen after the 
surface matter becomes optically thin. 

\subsection{Directional Dependence of Ejecta}

In the previous section we have examined the angular
averaged ejecta velocity composition and we show
that the ejecta composition depends on
the explosion geometry. Here we further on
analyze the ejecta composition by choosing specific
angular slices. In particular we choose 
the angular slices at 0 -- 9 $\deg$ from the 
rotation axis and from the symmetry axis (i.e. 81 -- 90 $\deg$
from the rotation axis)
to contrast the ejecta composition. This demonstrates 
the difference between the metal production with or
without geometric convergence.

\begin{figure*}
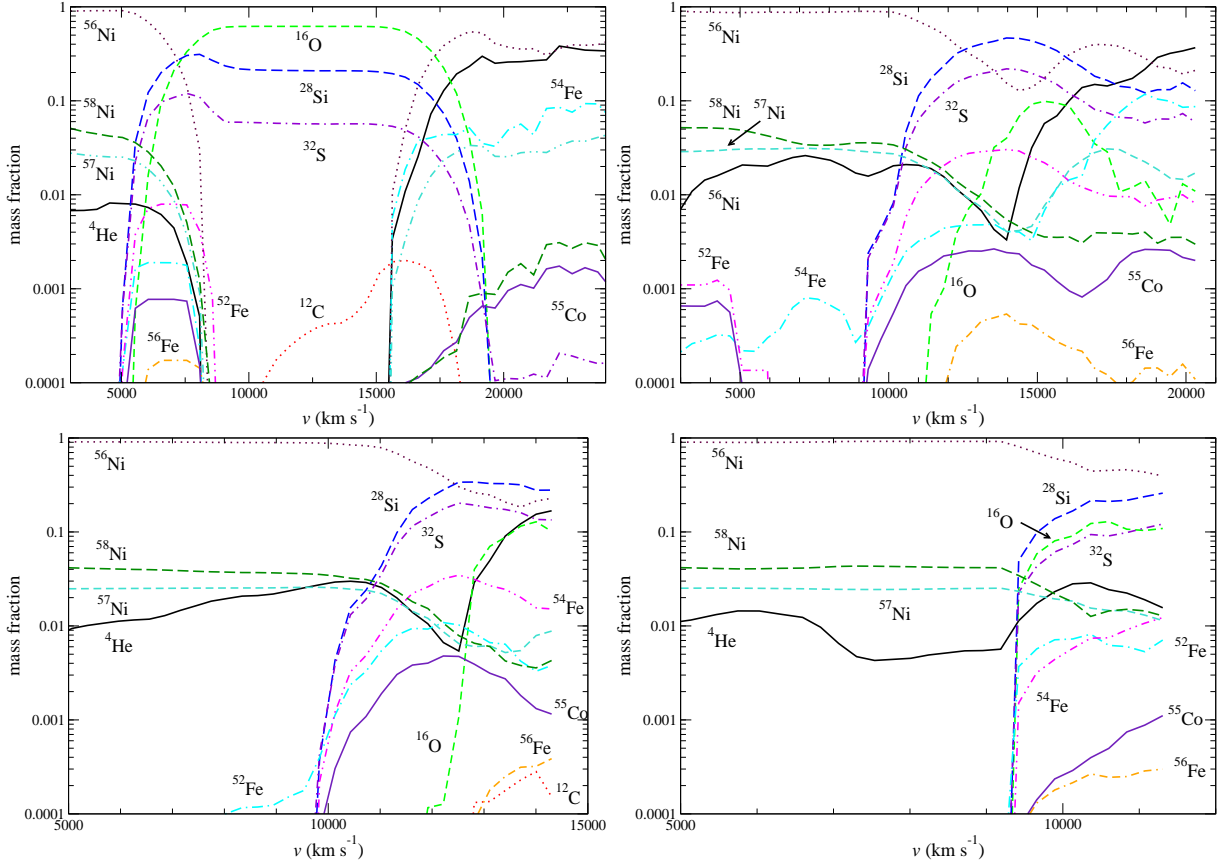

\centering

\includegraphics*[width=8cm,height=5.7cm]{fig5a.eps}
\includegraphics*[width=8cm,height=5.7cm]{fig5b.eps}
\includegraphics*[width=8cm,height=5.7cm]{fig5c.eps}
\includegraphics*[width=8cm,height=5.7cm]{fig5d.eps}
\caption{(top left panel) Ejecta compositions in the velocity
space for Model 100-100-B along the angular slice from 0 -- 9 $\deg$
from the rotation axis. (top right panel) Same as the top left
panel, but for the angular slice 81 -- 90 $\deg$ from
the rotation axis. (bottom left panel) 
Ejecta compositions in the velocity space for Model
100-100-R along the angular slice from 0 -- 9 $\deg$
from the rotation axis. (bottom right panel) Same as the
bottom left panel, but for the angular slice 81 -- 90 $\deg$
from the rotation axis.}
\label{fig:vel_map_slice}
\end{figure*}

\subsubsection{Bubble-Type Explosion}

The one-bubble configuration has shown to be weaker than the 
one-ring counterpart due to the lack of shock convergence. 
The explosion is weaker with a lower $^{56}$Ni production.
In the top left and right panels of Figure \ref{fig:vel_map_slice} 
we plot the ejecta distribution in the velocity space 
of Model 100-100-B for 
two angular slices, near the rotation axis (left) and near the symmetry
axis (right).

The two slices show very different ejecta structures. 
The ejecta along the rotation axis is slightly faster than 
the rotation axis. It is because the initial detonation triggered
along the "poles". However we remark that it does not mean 
there is more substances along the axis because the velocity
space does not follow the mass coordinate directly.
The iron-peak elements in the core also differ a lot.
Along the "poles", the representatives of iron-peak elements including 
$^{56,57,58}$Ni occupy the innermost $\sim 7000$ km s$^{-1}$
and then the abundance quickly drops off. On the other hand,
the iron-peak element-rich core extends up to 12000 km s$^{-1}$.
along the "equator". The large difference comes from the 
C-detonation. It is triggered near the "equator". As a result, 
the C-detonation first burns the material along the 
"equator" and then reaches the center, and then burns the matter
along the "equator". There is more time for the matter to move
outwards and expands, thus yielding a weaker heating effect.

The IMEs form the middle layer from 7000 -- 17000 km s$^{-1}$
along the "poles" and from 12000 -- 15000 km s $^{-1}$
along the "equator". Along the "poles", as the velocity increases,
which corresponds to lower density matter, some unburnt $^{12}$C
can be seen. However, there is no such trace along the "equator". 
Again, this demonstrates that the detonation along the 
"poles" is weaker than that along the "equator" due to the time lapse
during expansion.

From 17000 -- 21000 km s$^{-1}$ along the "poles" $^{56}$Ni dominates
the ejecta again. These are the product of the He-detonation
as the $^{4}$He mass fraction along becomes significant. They 
share similar mass fractions up to the surface. On the other
hand, along the "equator" there is more $^{56}$Ni 
from 15000 -- 19000 km s$^{-1}$
and more $^{4}$He from 19000 km s$^{-1}$ onward.

\subsubsection{Ring-Type Explosion}

The one-ring configuration on the other hand shows stronger
explosion than the one-ring counterpart through the shock convergence
near the "poles". The explosion is also stronger with 
a higher $^{56}$Ni production. 
In the bottom left and right panels of Figure \ref{fig:vel_map_slice}
we plot the same as the top panels but for Model 100-100-R.

Along the 
"poles",
the $^{56}$Ni builds the core of iron-peak elements which extends up to 12000 km s$^{-1}$.
Other iron-peak elements such as $^{57}$Ni and $^{58}$Ni are overwhelmed
by the IMEs ($^{28}$Si and $^{32}$S as two representatives)
at a lower velocity of 10000 km s$^{-1}$. 
Beyond 12000 km s$^{-1}$, IMEs are the dominant species until the surface.
A minor jump of $^{56}$Ni can be seen only near the surface,
unlike Model 100-100-B. $^{4}$He is also much lower than 
the IMEs, and is $\sim 20 \%$ of the surface abundance. 

On the other hand, there is no such transition of $^{56}$Ni-rich core
to IME-envelope along the "equator". A rapid jump of IMEs appears 
near 9000 km s$^{-1}$. However, the total abundance is about 0.1 -- 0.2
lower than $^{56}$Ni.

\subsubsection{Model Comparison}

The models 100-100-B and 100-100-R constitute two extremes 
of initial detonation configurations with the minimal perturbation 
from the spherical detonation. 
The distribution of elements in the velocity space appears to be very distinctive.

Without sufficient shock convergence, the explosion ejecta
consists of much stronger traces of IMEs in the middle
layer and $^{56}$Ni-$^{4}$He transition near the surface.
The distribution of $^{56}$Ni is discontinuous along some direction
in the "B"-Type model, but continuous in the "R"-Type model.
Meanwhile, IMEs are more pronounced in the "R"-Type model 
near the surface but not the "B"-Type model.
These distinctive features can be the indicator of where
the initial He-detonation starts.

\subsection{$^{56}$Ni Mass: Dependencies on the WD Mass and Detonation Morphology}

\begin{figure}
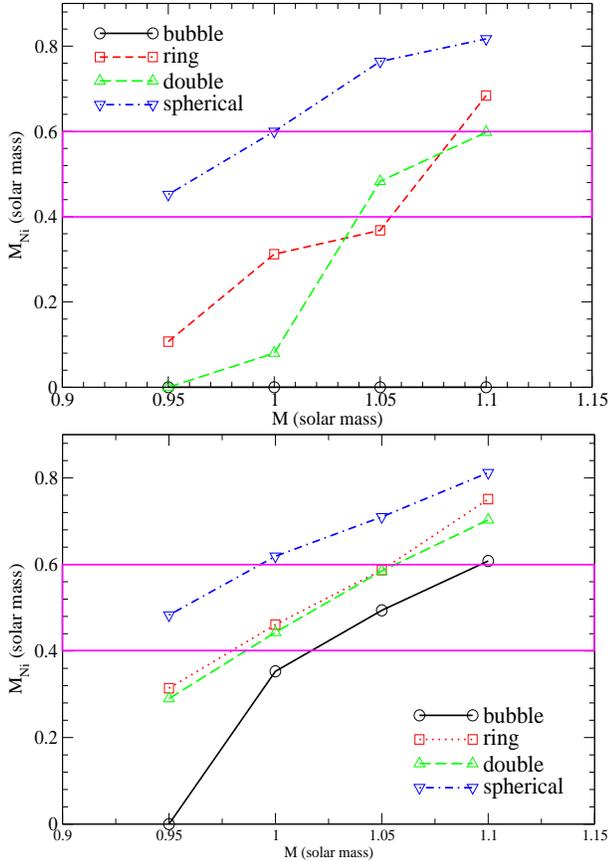

\centering
\includegraphics*[width=8cm,height=5.7cm]{fig7a.eps}
\includegraphics*[width=8cm,height=5.7cm]{fig7b.eps}
\caption{$^{56}$Ni mass against $M$ for the four
different initial He-detonation configurations for 
$M_{{\rm He}} = 0.05 ~M_{\odot}$ (upper panel) and 
0.10 $M_{\odot}$ (lower). The magenta box represents the  
range of $^{56}$Ni derived from the 
gamma-ray data from SN 2014J (\S \ref{sec:SN2014J}).}
\label{fig:Mni_M}
\end{figure}

The explosion of a sub-Chandrasekhar mass WD is known to be sensitive to 
the initial WD mass, $M$. It is because the WD is degenerate,
so that the central density varies largely from $10^{7}$ g cm$^{-3}$
(for $M \sim 0.95 ~M_{\odot}$) to $10^9$ g cm$^{-3}$ 
(for $M \sim 1.20 ~M_{\odot}$). The difference in the central density
corresponds to a difference in the average density of WD matter,
where the CO matter can be completely burnt to $^{56}$Ni
when it has a typical density $\sim 5 \times 10^7$ g cm$^{-3}$.
As a result, the amount of final $^{56}$Ni drastically varies
by a factor of 2 -- 6 when $M$ increases from 0.9 to 1.2 $M_{\odot}$.

In Figure \ref{fig:Mni_M}, we show the final $^{56}$Ni mass against
$M$ for the four different detonation morphologies described in the previous
subsections. 
Here the magenta box represents the range of $^{56}$Ni derived from
the gamma-ray data from SN 2014J.  See \S \ref{sec:SN2014J} for the
application to SN 2014J.
Models in which the C-detonation cannot be triggered are omitted,
i.e. 095-050-B, 095-050-B. 100-050-B, 105-050-B,
110-050-B, and 095-050-D. All of these show a typical He nova
event without the C-detonation \citep{Kippenhahn1978,Piro2004}, 
which is inconsistent with the 
supernova observation that the whole star is disrupted after the 
explosion. 
From both figures, we can see that the "S"-series is the
strongest, and then "D"- and "R"-series. The "B"-series is 
the weakest for the same initial $M$.

The possibility of triggering the C-detonation in the low $M_{{\rm He}}$ 
limit relies on the detonation symmetry.  The CO WD
models with a lower mass from $M = 0.95$ to $1.10 ~M_{\odot}$
show that, with the 
lowest symmetry ("B"-type), there is no geometric convergence. 
The only shock collision occurs when the detonation reaches
the "equator" of the WD. The assumed boundary condition (reflection symmetry)
allows the arriving shock
waves to collide in a laminar way. 
As indicated in Paper II, when $M$ increases, 
the minimum $M_{\rm He}$ required for triggering the second detonation
decreases. According to the similar work without assuming reflection symmetry \citep[e.g.,][]{Fink2014},
the minimum necessary $M_{\rm He}$ drops 
from 0.126 to 0.0035 $M_{\odot}$ when the 
CO WD mass increases from 0.810 up to 1.385 $M_{\odot}$.
Both works give us an insight that a higher $M_{{\rm He}}$
is necessary to provide the sufficient shock strength
in triggering the C-detonation.

By only considering the white dwarf mass, a higher $M$ 
means that the transition from the CO core to 
the He envelopes takes place at a higher density. 
This increases the typical reaction rate and hence 
the energy production. 
The post-shock temperature in the
He-envelope is, therefore, higher for more massive white dwarfs,
where the burnt matter can reach the threshold temperature
easier, independent of additional geometrical convergence. Thus, 
a higher $M$ model favours the trigger of the second detonation.

We remind that, in simulations using a hemisphere of a WD, 
the dependence on the $M_{\rm He}$ is stronger than 
simulations using a quadrant. It is because the detonation starts from
one pole and then the detonation wave wraps over the 
He-envelope and converges at the other pole. In this
situation a shock convergence similar to the "R"-Type detonation
always happens. As shown 
in the table for the "R"-Type detonation, the corresponding 
minimum $M_{{\rm He}}$ for second detonation is lower.

For "D", "R" and "S", they have a higher symmetry where
there is geometric convergence by means of 
oblique shock, two-dimensional shock convergence (from a ring to a point)
and three-dimensional shock convergence (from a sphere to a point) respectively. 
The resultant temperature in the CO core can be much
enhanced by the converged shock strength. 
Therefore, the minimum $M_{{\rm He}}$ required
to trigger the second detonation is more relaxed. 

The difference in the final $^{56}$Ni mass for the same $M$
at different He-detonation is related to the propagation 
of the C-detonation direction. The "S" model is 
always the strongest because the C-detonation begins 
at the center and propagates outward, so that most 
of the star remains approximately static before the detonation
wave arrives. This ensures the matter remains less expanded
and hence maintaining a higher density, which results in a stronger 
explosion. On the contrary, in the "B", "D" and "R" models,
the off-center C-detonation means that the C-detonation has
to overcome the density gradient in order to reach 
the high density matter in the core. This implies that the 
relative explosion strength is weaker because of the 
density gradient.

\subsection{Morphology of Remnants}

\begin{figure*}
\centering
\includegraphics*[width=20cm,height=5.7cm]{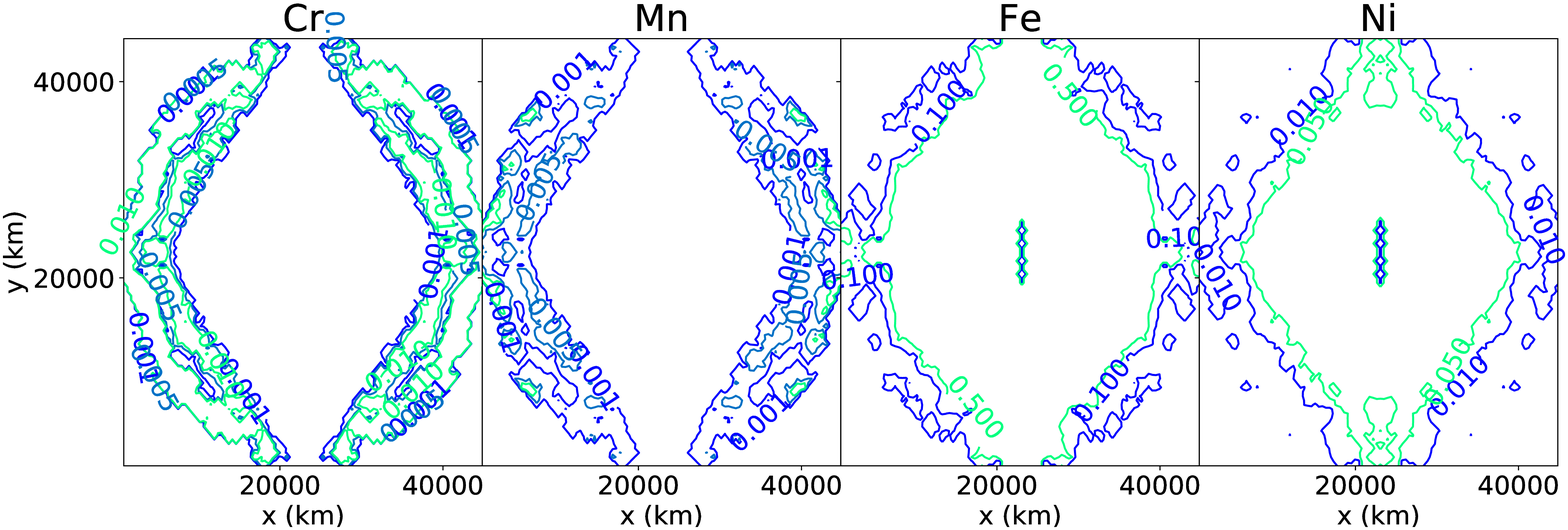}
\includegraphics*[width=20cm,height=5.7cm]{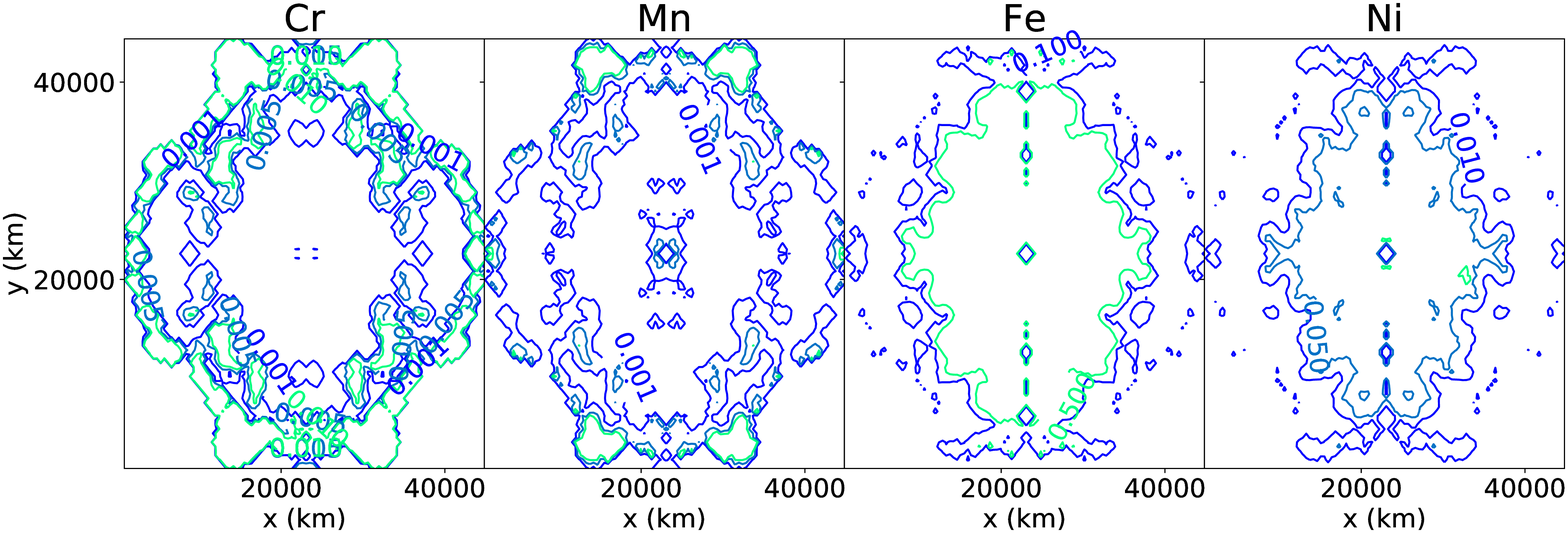}
\includegraphics*[width=20cm,height=5.7cm]{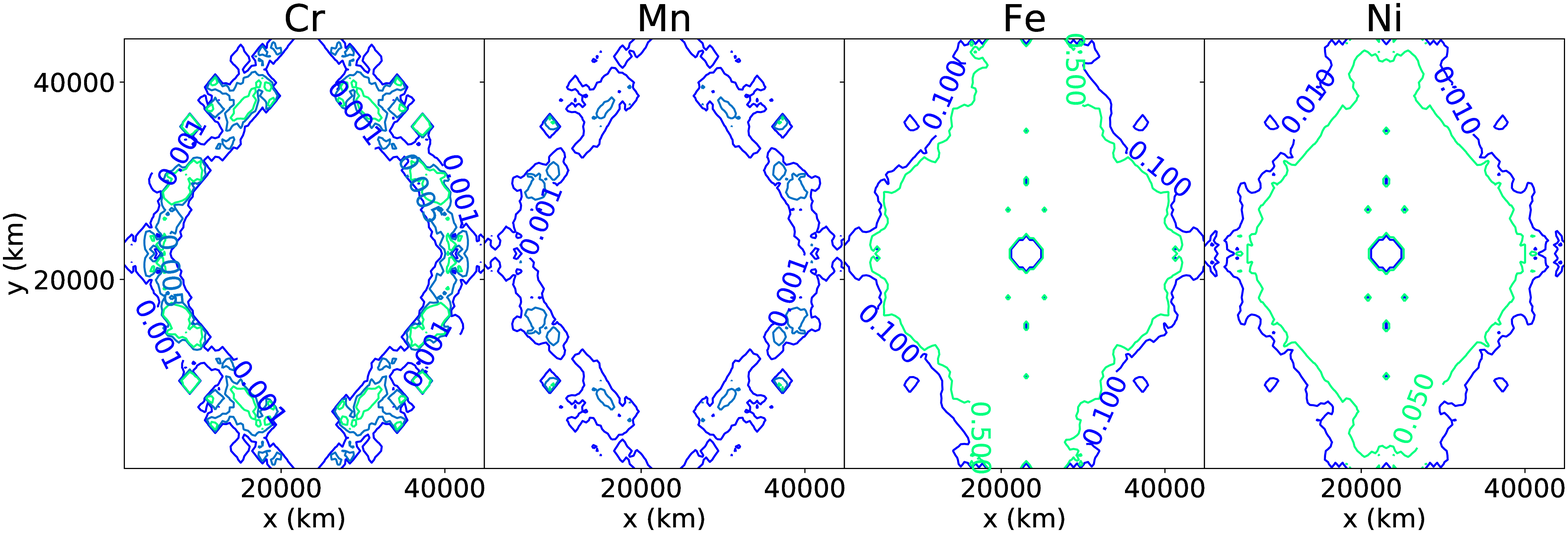}
\includegraphics*[width=20cm,height=5.7cm]{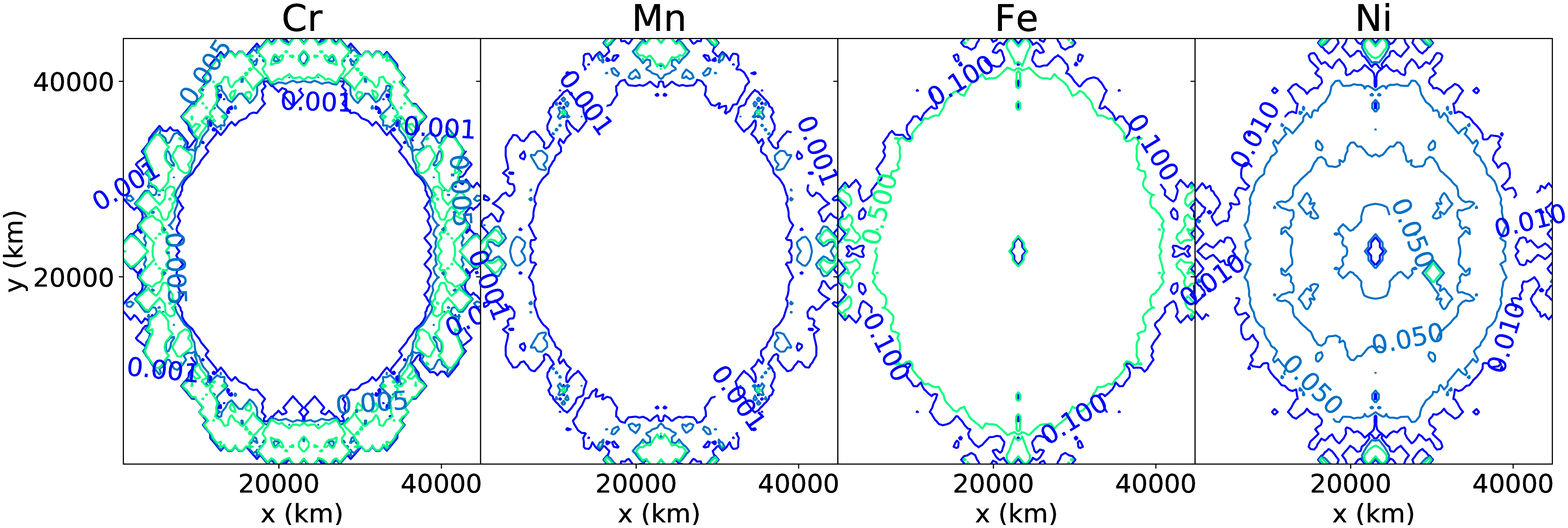}
\caption{Stable element mass fraction distributions of the 
explosion models 100-100-B ($1^{\rm st}$ row), 100-100-R ($2^{\rm nd}$ row), 
100-100-D ($3^{\rm rd}$ row) and 100-100-S ($4^{\rm th}$ row)
for Cr, Mn, Fe and Ni for the panels from left to right.
The numbers stand for the corresponding  
mass fraction contours for these isotopes.}
\label{fig:EleDist}
\end{figure*}

The early $^{56}$Ni in the ejecta near the surface provides 
distinctive hints on the asphericity of the detonation model. 
Furthermore, the distribution of iron-peak elements, as later they become
heat-shocked by the backward shock in the ejecta, 
they can reach a temperature $\sim 10^6$ K, which is sufficient
hot for X-ray emission for further diagnosis. This
will give further constraints on the explosion models. 
For example, the X-ray spectra of the SN remnant 3C 397
has been used to a diagnosis of a Chandrasekhar mass WD
progenitor \citep{Yamaguchi2015,Leung2018Chand,Dave2017}.

In Figure \ref{fig:EleDist} we plot the 
element distribution of Cr, Mn, Fe and Ni for the Models 
100-100-B, 100-100-R, 100-100-D and 100-100-S. We assume that 
after the star reaches homologous expansion the distribution of 
ejecta remains mostly unchanged. However, we also remark that 
the radiation energy during decay of radioactive isotopes can 
still trigger inner motion and affect the element distribution.
But the effect is secondary compared to the initial distribution
during explosion.

The distribution of the iron-peak elements show more diversity 
than the major elements as shown in the previous section. From the 
contour shape we observe two features. First, Cr and Mn almost 
follow each other. And Fe and Ni follow each other. Second, Cr and Mn 
tends to have a more spherical distribution while Fe and Ni 
follow more closely to the detonation geometry. 

Model 100-100-R shows the the largest deviation from a spherical structure among all four
elements. Model 100-100-B has spherical Cr and Mn but aspherical
Fe and Ni. Similar characteristics appear for Model 100-100-D. All elements
are spherical in the Model 100-100-S, as anticipated by the initial
spherical symmetry. To disentangle between Model 100-100-B and 100-100-D, 
we notice that the distribution of Cr is more irregular in Model 100-100-D,
compared to the quasi-spherical distribution in Model 100-100-B. 
However, the difference is subtle. 

\section{Case Study: Application to SN 2014J}
\label{sec:SN2014J}

\subsection{The Inspiring Case of SN2014J}

SN 2014J is a special example of SNe Ia exploded in the nearby galaxy
M82 just 3.3 Mpc away, the closest SN Ia in the last four decades. Its
vicinity to from the Milky Way galaxy has provided the chance for
detailed multi-band observations including the radio
\citep{PerezTorres2014}, infrared, optical \citep{Goobar2014,
Kawabata2014}, UV \citep{Foley2014}, X-ray \citep{Terada2016} and
gamma-ray \citep{Siegert2015, Diehl2014, Diehl2015a, Diehl2015b, Churazov2015, Isern2016} bands, with its spectra at early and late
time \citep{Ashall2014, Jack2015, Dhawan2018}.

Measurements of SN 2014J have been made in different works in the literature. 
In \cite{Churazov2014}, the estimated $M_{\rm ^{56}Ni} = 0.56 \pm ^{0.14}_{0.06} ~M_{\odot}$
and the estimated ejected mass is $1.2 \pm ^{1.9}_{0.5} ~M_{\odot}$.
A similar measurement is found in \cite{Diehl2015a} which gives
$M_{\rm ^{56}Ni} = 0.50 \pm 0.12 ~M_{\odot}$.
For $^{57}$Ni, \cite{Yang2018} reports that $^{57}$Ni/$^{56}$Ni
has a mass ratio 0.065$\pm ^{0.005}_{0.004}$ based on the B-band maximum light
and 0.066$\pm ^{0.009}_{0.008}$ based on the pseudo bolometric light curve. 
Stable Ni mass is constrained at $0.053 \pm 0.018 ~M_{\odot}$ 
\citep{Dhawan2018}.

It shows gamma-ray features which agree well with the classical spherical 
pure deflagration model W7 \citep{Churazov2014,Diehl2015a}.
In photometry, SN 2014J appears to be a normal SN Ia \citep{Isern2016} which shows 
a comparable structure with the W7 model \citep{Nomoto1984},
and a similar abundance profile with heavy elements (e.g., $^{56}$Ni)
in the core and lighter elements (e.g. Si and S) in the envelope
\citep{Ashall2014}.

However, detailed examinations of the observational data of SN 2014J
reveal some differences from ordinary SNe Ia.  For example, the rise
of UVOIR light curve with time \cite[e.g.][]{Nugent2011} shows its
delay in SN 2014J. The early light curve of SN 2014J suggests a
`shoulder' only a few days after the first light \citep{Goobar2014}.
The late time evolution (beyond few hundred days) shows derivations
from classical SNe Ia such as SN 2011fe, where the slower decline rate
suggests interactions with CSM \citep{Foley2014,Yang2018}.  The
ultraviolet data of SN 2014J shows large extinction \citep{Brown2015}.
Such extinction and CSM can be in the dusk disk structure
\citep{Nagao2017}.  Images around SN 2014J do not show an observable
companion star, thus making its companion as a red giant unlikely
\citep{Margutti2014,PerezTorres2014,Kelly2014}.  

It seems that the progenitor of SN 2014J is still a question of debate
\citep[e.g.,][]{Margutti2014,PerezTorres2014,Dragulin2016,Graur2019}.
We note that these constraints on CSM cannot be applied to the
presupernova environment of a uniformly rotating white dwarf with a
slightly super-Chandrasekhar mass in the single degenerate scenario as
calculated by \cite{Benvenuto2015}.  The gamma-ray signal cannot
distinguish with high significance which class SN 2014J belongs to
\citep{Terada2016}.

\subsection{Aspherical Features of SN 2014J and Constraints on Models}

According to further examinations, the observational data of SN 2014J
show features which deviate from the spherical approximation as
discussed below.

First, the early gamma-ray observations with INTEGRAL discovered the
lines at 158 and 812\,keV ($T_{1/2} = 6.6$\,d) that are characteristic
for the $^{56}$Ni decay around 17.5 days after the inferred explosion
date \citep{Diehl2014,Isern2016}.
The model fits of \cite{Isern2016} appeared more consistent 
with a red-shifted and broadened $^{56}$Ni emission, so that they suggested 
ejection of $^{56}$Ni-rich material in a blob moving away from the observer. 
The analysis by \cite{Diehl2014} was performed in finer energy bins, and 
without any model bias; their sampling of possible spectral solutions
suggest a narrow emission line from $^{56}$Ni at the laboratory energy value, 
with indications of blue- as well as red-shifted emission lines.
This led them to suggest a model with $^{56}$Ni 
ejected perpendicular to the observer's line of sight. But the total 
significance of this surface $^{56}$Ni line emission is only 3 -- 4
sigma, hence both interpretations remain possible.

The second aspherical feature appears at later times.  When the energy
output is dominated by the decay of $^{56}$Co ($T_{1/2} = 77.1$\,d),
the measured Doppler-shifts of the $^{56}$Co decay lines show the expected
behavior, plus an additional structure of (at least) three blobs of
distinctive velocities \citep[i.e. showing an early blue-shift when
the red-shifted part is opaque in the line of sight, and then
becoming symmetric with no Doppler-shift later as pointed out
in][]{Diehl2015a}.  Instead, at least three distinctive centroid
energies could be identified. This may correspond to fluid parcels
containing $^{56}$Co are ejected with different velocities with
respect to the Earth frame.  This time-dependent variations of the Co
decay line frequency, denoted as flickering, is 
observed in SNe Ia for the first time.  This suggests the possibility that `blobs', or
large scale asymmetries, developed during the explosion.

To summarize, SN 2014J has shown that the theoretical conflicts
with the classical spherical model.  These include:
(1) the observed Doppler-broadened $^{56}$Co lines, which have
an irregular appearance with time \citep{Diehl2015a}, 
(2) hints of the $^{56}$Ni decay lines on the surface \citep{Diehl2014}, and
(3) an enhanced ionization on the outer part of the star
as revealed by the early atomic line spectra.

\subsection{Near-Chandrasekhar Mass Models That Produce Surface $^{56}$Ni}

Among the above mentioned features of SN 2014J, the existence of
$^{56}$Ni near the surface suggested by the INTEGRAL data is the main
motivation for the present study.  The missing of the C and O
absorption lines is possibly connected to the exposed $^{56}$Ni lines
\citep{Goobar2014}.  It might be worth noting that these expected Ni
lines are not seen in infrared at this time.

Before discussing the sub-Chandrasekhar mass models in the following
subsections, we note that near-Chandrasekhar mass models have
variations depending mainly on the mass accretion rate, and some
models produce $^{56}$Ni near the surface of WDs.  In the classical
picture of the near-Chandrasekhar mass model, $^{56}$Ni is
concentrated in the inner core \citep{The2014}.  However, $^{56}$Ni is
produced near the surface in the following models.

\cite{Nomoto1982b} showed several models where the WD mass increases
to the near-Chandrasekhar mass with slow accretion of He.  If the
accretion rate is low enough, the accreted He is too cold to be
ignited, thus being just accumulated near the surface.  Eventually,
the WD becomes massive enough to ignite the central C-deflagration.

In the {\sl late} detonation model by \cite{Yamaoka1992}, the
deflagration-detonation transition can occur in the outer layer of the
near-Chandrasekhar mass WD and burns He to produce $^{56}$Ni near the
surface \citep[see model W7DHE in Fig. 3 of][]{Yamaoka1992}.  Such a
deflagration-detonation transition might occur at the very steep
density gradient near the WD surface even with a small amount of He.
The deflagration-detonation transition near the surface would be likely
to occur in an aspherical manner which might be interesting for
further study.

\subsection{Modeling Issues on Sub-Chandrasekhar Mass Models}

Although the production of $^{56}$Ni near the surface is possible in
the late detonation of the near-Chandrasekhar mass model (W7DHE), we
here focus on the 
sub-Chandrasekhar mass models in order to
apply the results of this Paper IV to the ``asphericity'' of SN 2014J.

SN 2014J was first proposed to be the ignition of a ``He belt'' accumulated 
in the orbital plane of the binary \citep{Diehl2014}.  The He belt model tried to 
explain the early $^{56}$Ni decay line detected with 
close-to-zero line Doppler shift.
This picture can explain the origin of the early time 
$^{56}$Ni decay line. However, on top of that, in \cite{Diehl2014}
such decay line has a very small redshift. 
The He belt model, when 
observed from the "north-" or "south-poles"\footnote{notice that 
in general a static WD is considered and the "poles" has a 
graphic meaning of being the upper and lower ends of the sphere,
while the "equator" means the symmetry plane between the two 
"poles".}, 
the ejection of Ni from the He envelope will be all along the
"equator" direction.  As a result, it provides the source of Ni with
small Doppler shift.  Also it remains unclear if the He belt can be
formed and maintained stably during the binary accretion
\citep{Kippenhahn1978,Piro2004}.

In the following discussion, 
we search for a qualitative
model that may resemble with the observed characteristics of SN 2014J.
We note that the model asphericity should not be too strong, as otherwise
it would violate the indicated proximity of SN 2014J with the
classical spherical model.
Thus we set the He envelope in spherical form.
To generate the required surface Ni, different
He-detonation configurations are studied.

Our attempt is to elucidate 
the physical conditions which are able to reproduce the distinctive 
features of SN2014J qualitatively. Owing to the complexity of multiple 
data constraints and the subtlety in the interpretations of different
observations, we avoid to scrutinize an exact or complete model that 
can explain all quantitative features for SN 2014J. Our focus is, therefore, 
on the early $^{56}$Ni line emission, the $^{56}$Ni mass as derived
from the peak (V-band) luminosity, large-scale asymmetries as implied
by the flickering $^{56}$Co decay line profiles, and the
$^{57}$Ni/$^{56}$Ni mass fraction ratio from the late-time light
curve. Using $^{57}$Ni/$^{56}$Ni to constrain the SN explosion models
as proposed in \citep{Seitenzahl2009} has been used in other SNe Ia,
for example SN 2012cg.

To recapitulate, based on the two-dimensional models, we aim at:
\noindent (1) searching for WD parameters which corresponds to the
general features of SN 2014J,
\noindent (2) searching for appropriate He-detonation triggers which
produce global asymmetries and qualitative features of SN 2014J,
including large scale asymmetry and/or near-surface production of
$^{56}$Ni.

\subsection{Constraints on Progenitor WD Mass}

The $^{56}$Ni mass observed in SN 2014J provides the important
constraint on the progenitor WD mass.
In \cite{Diehl2015a}\footnote{Other works in the literature 
give a similar range. For example in \cite{Churazov2014}
the upper limit can reach 0.7 $M_{\odot}$, but the 
uncertainties are similar.}, the $^{56}$Ni mass is estimated to be
$0.49 \pm 0.10 ~M_{\odot}$. From our simulation results in Table 1, we identify 
the possible mass range to be $M = 1.0 - 1.1 ~M_{\odot}$.

In Figure \ref{fig:Mni_M}, we show the final $^{56}$Ni mass against
$M$ for the four different detonation morphologies described in the previous
section.
To account for the observed $^{56}$Ni mass in SN 2014J shown by the
magenta box in Figure \ref{fig:Mni_M}, we require $M = 0.95 - 1.00
~M_{\odot}$ for the "S"-series and $M = 1.00 - 1.10 ~M_{\odot}$ for
the "B"-series ($M_{{\rm He}} = 0.10 ~M_{\odot}$).  "D"- and
"R"-series require $M = 1.05 - 1.10 (1.00 - 1.05) ~M_{\odot}$ for
$M_{{\rm He}} = 0.05 ~(0.10) ~M_{\odot}$.

\subsection{Constraints on Explosion Mechanisms}

Another aspect to constrain SN 2014J is by the explosion geometry.
The gamma-ray signal of SN 2014J has suggested the (near-)surface
$^{56}$Ni. We examine which initial detonation geometry
allows the formation of $^{56}$Ni at these regions. To extract
the final distribution of $^{56}$Ni, we use the tracer particle
data when the ejecta reaches homologous expansion and 
the structure of the ejecta is frozen out. 

Future observations of SN 2014J, which will be able to disentangle the morphology, 
may thus provide clues to the initial configuration from measurements
of different elements. For example, 
in \cite{Grefenstette2014,Grefenstette2017} such a technique is pioneered in 
showing $^{44}$Ti in Cas A to demonstrate how to 
disentangle and map the gamma-rays.

In Figure \ref{fig:NiDist}, we show the distributions
of some representative elements for models with different detonation geometry. 
For example, for models starting
with a He-detonation bubble, the Ni and Si ejecta are in a cocoon 
shape, compared to the spherical shape in the spherical detonation. 
The "B"-Type is also highly distinctive by the thick layer of the O-rich 
ejecta, compared to the spherical counterpart. On the other hand, 
"D"-Type and "R"-Type are different from the other two by 
the quasi-spherical O- and Si-rich ejecta, while the Ni-ejecta
maintain an observable ellipticity.
However, between D and R type models, Si and Ni show a similar distribution.

\subsection{Constraints on the He Envelope Mass}

\begin{figure*}
\centering
\includegraphics*[width=8cm,height=5.7cm]{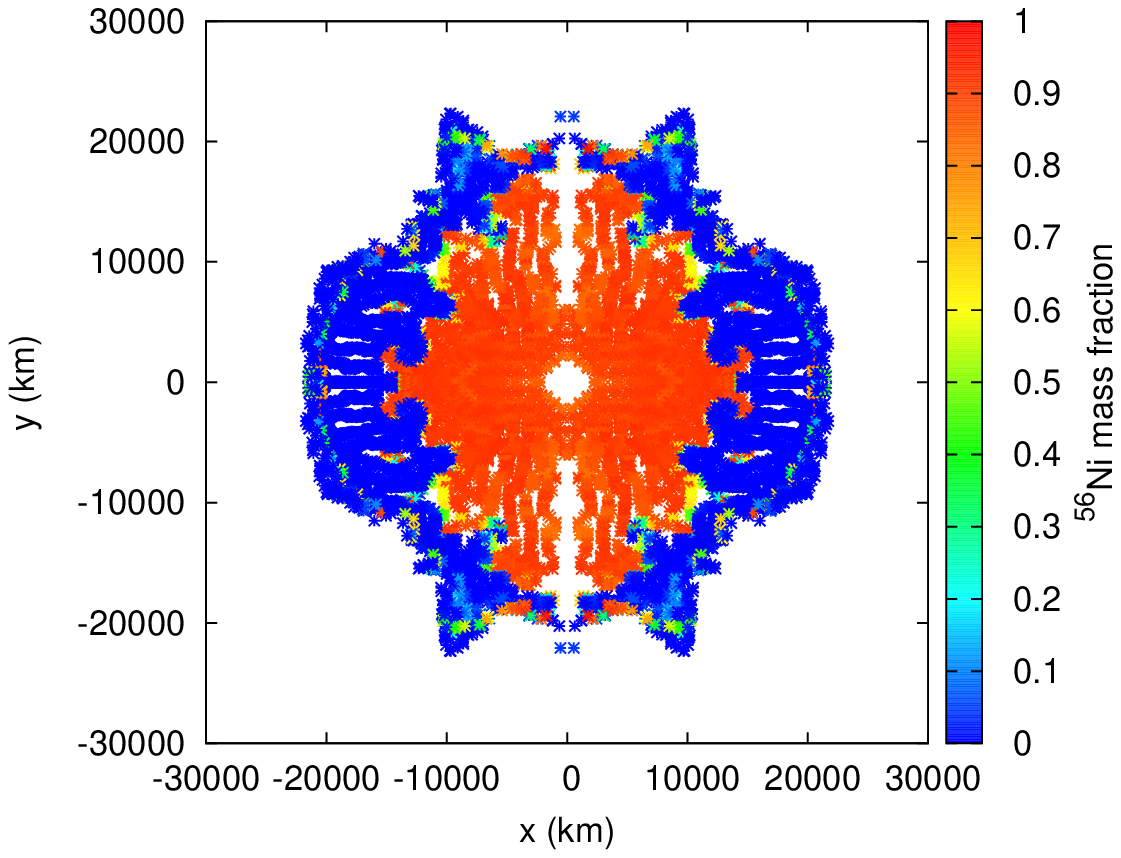}
\includegraphics*[width=8cm,height=5.7cm]{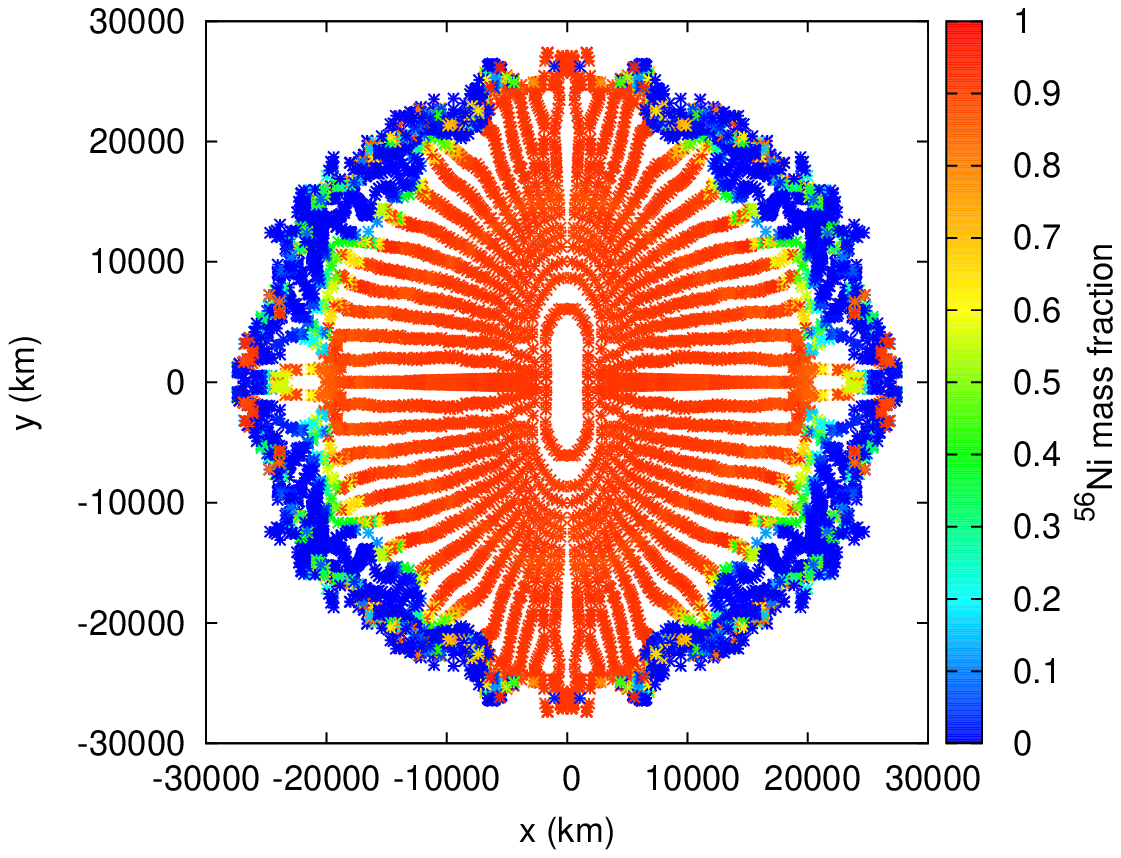}
\caption{
The $^{56}$Ni mass fraction distribution of the 
explosion models 100-050-R (left panel) and 100-100-R (right panel).}
\label{fig:NiDist_He}
\end{figure*}

We now examine the dependence of the large scale asymmetry on the He
envelope mass.  In Paper II we have presented a parameter survey on
presented a parameter survey on the nucleosynthesis yield 
of SNe Ia using the sub-Chandrasekhar mass WD as the 
initial progenitor. It is shown that the He-envelope
mass can strongly enhance the production of some
Fe-peak isotopes, including $^{48}$Ti, $^{50,51}$V
and $^{52}$Cr.

In Figure \ref{fig:NiDist_He}, we plot the $^{56}$Ni distribution for
Models 100-050-R and 100-100-R. They differ by the 
mass of the He-envelope from 0.05 to 0.1 $M_{\odot}$. 
The distribution of $^{56}$Ni after the explosion can 
show the existence of a large-scale asymmetry. 
In Model 100-100-R, with a more massive He envelope, 
the He-detonation is strong enough to drive the inwards propagation of the detonation
to the core; this results in the complete disruption, 
where most of the matter in the CO core is spherical.

Therefore, this scenario will less likely exhibit a flickering as seen in SN 2014J. 
However, its $^{56}$Ni distributes also close to the surface,
which is an important feature of SN 2014J.
As a result, the ejected $^{56}$Ni around 10 degrees from the 
rotation axis has a higher radial velocity. The 
lower He-envelope mass makes the production of $^{56}$Ni 
lower, and the synthesized $^{56}$Ni is covered by the original
He-envelope, which can block the gamma-rays emitted by 
the radioactive decay. This means a lower $M_{\rm He}$ model 
has more difficulties to reproduce the early gamma-rays observed from SN2014J.

Following the expansion of the ejecta, when the 
matter becomes optically thin to the gamma-ray, 
the inner structure of the $^{56}$Ni and $^{56}$Co 
distribution will be exposed. The velocity fluctuation,
as seen from Figure \ref{fig:NiDist_He},
depending on the ejecta angle and their exact time
to become optically thin, may coincide with the 
flickering feature as seen in SN 2014J.

\subsection{Constraints from Nucleosynthesis}

\begin{figure}
\centering
\includegraphics*[width=8cm,height=5.7cm]{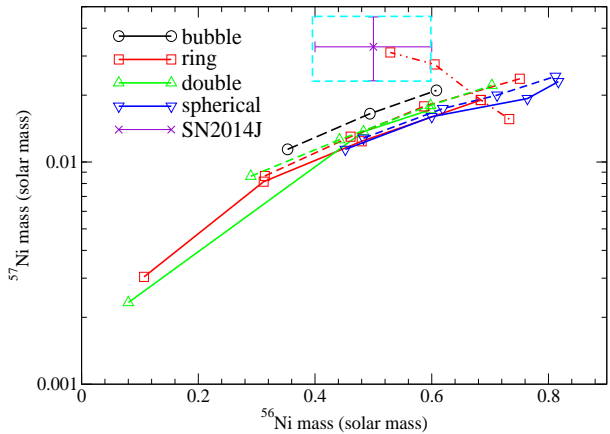}

\caption{The $^{57}$Ni mass against $^{56}$Ni for the four
different initial He-detonation configurations for 
$M_{{\rm He}} = 0.05 ~M_{\odot}$ (solid line) and 
0.10 $M_{\odot}$ (dashed line) and for solar metallicity. The data point 
shows the observed $^{57}$Ni/$^{56}$Ni mass ratio
derived from the late-time light curve of the 
optical band. The red dash-dot-dot line corresponds
to the sequence of Model 110-050-R50 using
different initial metallicity of $Z = 0, 1, 3$
and 5 $Z_{\odot}$ (from bottom to top). The
data points from left to right corresponds to the models
with $M = 0.9$, 1.0, 1.05 and 1.1 $M_{\odot}$.}

\label{fig:Ni57_Ni56}
\end{figure}

In Figure \ref{fig:Ni57_Ni56} we plot the $^{57}$Ni 
and $^{56}$Ni yields (prior to decay) of our simulated 
SN Ia models. 
Shown are models with $M_{\rm He} =0.05$ (solid lines) 
and $M_{\rm He} = 0.10~M_{\odot}$ (dashed lines) at solar metallicity. 
We also 
show the SN 2014J data by the cross symbol \citep{Yang2018}. 
Apparently the models listed there are not sufficient to explain the 
high $^{57}$Ni mass of this supernova.

Similarly, to demonstrate the effects of metallicity, we plot
the dash-dotted line of Model 110-050-B 
for $Z =$ 0, 1, 3 and 5 $Z_{\odot}$ (corresponding to the data point
from the bottom to the top). To explain the 
SN 2014J data, a model with $Z \sim 5~Z_{\odot}$ is
required. All the trend lines for other models
behave similarly. In Paper I we have
shown that the metallicity has minor impact to the 
global explosion energetic. Instead, it shows its influence
on the relative abundance ratio, in particular the 
high-Y$_e$ isotopes including $^{54}$Fe, $^{58}$Ni
and $^{55}$Mn. 

From this figure we observe that in order to explain the 
abundance pattern of SN 2014J, a high metallicity model with
$Z \gtrsim 4 ~Z_{\odot}$ is necessary to be consistent with the
high $^{57}$Ni mass relative to $^{56}$Ni.

\section{Extension to Other Supernova Observations}
\label{sec:extension}

\subsection{Remnant Morphology}

The morphology of the supernova ejecta and the shape of 
the SN remnant, such as Tycho \citep{Gilles2019}, may directly link to the initial
explosion configuration.
By tracing the line emission of shock heated ejecta, 
the abundance of the measured elements can be revealed.
In \cite{Seitenzahl2019} the tomography of three youngest SN Ia remnants,
0519-69.0, 0509-67.5 and N103B in the large Magellanic Cloud
\citep{Hughes1995} are studied for the first time. 
By examining the S-XII, Fe-IX and Fe-XV lines, they reconstructed
the large-scale distributions of these elements in these remnants. 
These objects are sufficiently young such that the shocked-heated
matter remains clearly visible and the shock front has not completely
swept all the matter.

SN remnant 0519-69.0 has a more spherical shape,
but with small scale perturbations on the surface as depicted
by the X-rays.  
From our simulations, such feature is possible when we 
consider the Rayleigh-Taylor instabilities which freeze
out during the expansion of the ejecta. 
As discussed in \cite{Seitenzahl2019}, this remnant can be fitted
by the Chandrasekhar mass WD of mass 1.4 $M_{\odot}$.
Their total Fe-mass is about 0.4 $M_{\odot}$. This belongs to 
the lower side of SN Ia production \citep{Leung2018Chand}.
Such features can be obtained for a higher mass progenitor
$\sim 1.4 ~M_{\odot}$ with an initial central density $> 3 \times 10^9$ g cm$^{-3}$.

SN remnant 0509-67.5 also has a spherical shape
where the ejecta including X-ray demonstrates a close-to-spherical
structure. The estimated mass of this remnant is 
$\sim 1.0 ~M_{\odot}$ with 0.15 $M_{\odot}$ He. 
To achieve such spherical shape, a spherical He-detonation
is necessary to avoid any large-scale asymmetry created
during the C-detonation. In fact, when such a heavy He-envelope
is included, spherical detonation is more preferred because
the expected nuclear runaway time can easily be shorter
than the convection time scale, thus triggering simultaneous
burning in the spherical layer near the He-CO interface.
Also, as discussed in Paper II, 
the thick He envelope $> 0.1~M_{\odot}$ can bring
a very severe excess in light iron-peak elements such as Ti, V and Cr.
Future detection of these elements may provide 
further confirmation in this explosion picture. 
At last, its total Fe mass $\sim 0.5 ~M_{\odot}$
can be mapped consistently to our models such as 100-100-S.

SN remnant N103B also has an aspherical ejecta
shape. However, not much analysis of this object is reported
in that work. Despite that, from the morphology of ejecta,
an arrow or cone shape of Fe XIV distribution can be observed.
Such an aspherical shape with such a pointing effect can indicate
the focused shock in the CO core and its consecutive breakout,
as hinted from Figure \ref{fig:ChandvssubChand}.
In our work, due to the reflection symmetry, the 
cone shape is always paired on both sides. We expect that,
if we allow single bubble without assuming reflection
symmetry, a one-sided feature can be resulted as done 
in more general simulation such as \cite{Tanikawa2019,Gronow2020}.

\subsection{Remnant Element Abundance}

\begin{figure}
\centering
\includegraphics*[width=8cm,height=5.7cm]{fig10.eps}

\caption{$^{56}$Ni production for the sub-Chandrasekhar mass
models with initial spherical detonation structure for 
$M = 0.9$ o $1.2 ~M_{\odot}$ at $Z = 0$, 0.002, 0.01, 0.02 and 0.04.}

\label{fig:Ni56_Z_plot}
\end{figure}

\begin{figure}
\centering
\includegraphics*[width=8cm,height=5.7cm]{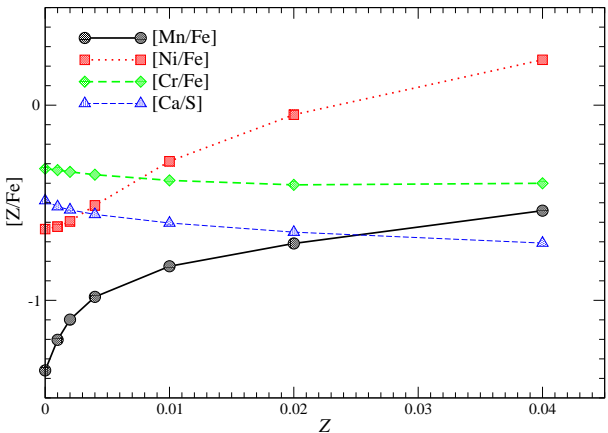}

\caption{Abundance ratios for the sub-Chandrasekhar mass
model with initial spherical detonation structure for 
$M = 1.0 ~M_{\odot}$ at $Z = 0$, 0.002, 0.01, 0.02 and 0.04.}

\label{fig:abundance_Z_plot}
\end{figure}

Another possibility to trace the remnant progenitor
is to examine the abundance patterns. As described in previous sections,
the X-ray spectra from SNR can provide essential clues 
on the relative amount of elements, especially iron-peak elements
including Cr, Mn, Fe and Ni \citep{Yamaguchi2015,MartinezRodriguez2017}.
They can directly constrain from which type of WD progenitor
and the explosion physics. In their works, the SNR spectra including 
Tycho, Kepler, 3C 397, and G337.2-0.7 in the Milky Way and N103B
in the Large Magellanic Cloud are analyzed. It is suggested that 
to explain the variety of remnant chemical abundance pattern,
factor beyond neutronization (i.e., metallicity tracer by $^{22}$Ne)
is necessary. Here we examine how the metallicity affects the 
elemental distribution and $^{56}$Ni. 

In Figure \ref{fig:Ni56_Z_plot} we plot the mass and $Z$ dependence
of $^{56}$Ni production for the sub-Chandrasekhar mass WD models
with $M = 0.9$ -- 1.2 $M_{\odot}$ and $Z = 0$ -- 0.04. 
The $^{56}$Ni-production is strongly $M$-dependent and 
monotonically increasing with $M$. The transition from 
complete burning to incomplete burning shows the strongest
effects between 0.9 -- 1.0 $M_{\odot}$ by a change of almost
a factor of ten within this mass range. The effects of metallicity
can be seen but is small compared to the effects of $M$. 
In general a $\sim 10 \%$ difference can be seen among 
the $Z$-range examined here. 

In Figure \ref{fig:abundance_Z_plot}, we plot the mass fraction ratio 
of [Mn/Fe], [Ni/Fe], [Cr/Fe] and [Ca/S] for our sub-Chandrasekhar
mass WD models with a mass 1.0 $M_{\odot}$ for metallicity 
$Z = 0$, 0.002, 0.01, 0.02 and 0.04. The metallicity effect
is much larger. Major IPES such as [Mn/Fe] and [Ni/Fe]
differ by $\sim 0.6$ dex between models with $Z = 0$ and $Z = 0.04$.
Both of them increase when $Z$ increases.
On the other hand, minor IPEs such as [Cr/Fe] and IMEs [Ca/S]
show much smaller variations by only 0.1 dex in the $Z$-range
examined. They decrease when $Z$ increases. 

However, we remark that the exact values of [Ca/S] and [Cr/Fe]
are more prone to systematic uncertainties. It is because 
the light IPEs including Ca and Cr can be produced in both 
NSE and incomplete-Si burning. For Mn, Fe and Ni, they are 
robustly produced in NSE matter, where complete and instantaneous 
energy release can be assumed. On the other hand, for Ca and 
Cr, the numerical scheme in how the low density burning 
can affect the formation of these elements. For the same 
amount of energy release, a longer energy deposition time can 
slow down the thermal expansion. As a result, the matter has
more time to carry out slow nuclear reaction in the 
$\alpha$-rich freezeout regime before the ejecta expands and 
becomes too cold for any significant nuclear reactions. 
The case for low density matter contains complication
because the actual reaction depends on the detailed chemical
composition, which is not well traced in multi-dimensional composition.
The simplified chemical composition (7-isotope network)
may not provide an accurate estimation in how fast those 
reactions and the associated energy production take place. 

\section{Discussion}
\label{sec:discussion}

\subsection{Connection to the Works in the Literature}
 
Our work has suggested that the He-detonation is the key feature
for explaining some well observed SNe Ia, such as SN 2014J. 
However, it also remains to be understood in details
how the ignition of surface He can be coupled to the current explosion mechanism
which ignites carbon centrally in the WD.
Three-dimensional Low-Mach number hydrodynamical simulations of the
He-burning envelope is necessary for realizing how the initial
hydrostatic He-burning develops into nuclear runaway.
The detonation size is typically 
assumed to be as large as the pressure scale height
\citep{Bildsten2007,Shen2009,Dan2014}. On the other hand, 
temperature fluctuations tend to favour the ignition
at a single spot, which has a much smaller size. 
A stringent limit on the temperature fluctuation
appears because of its small size \citep{Holcomb2013}.
Contamination of C-rich matter from 
the CO-core can be an alternative to decrease the necessary
size of nuclear burning \citep{Shen2014}. 

The pre-runaway phase of the double detonation model
is unclear until recent large scale works on clarifying 
the possibility of such proposal \citep{Jacobs2016}. 
Depending on the convection flow, 
different He-runaway pattern scenarios could occur from the most
non-spherical extreme, i.e. a bubble, to the most spherically
symmetric case. The modeling of 
such process is typically much longer than the hydrodynamics
timescale, in order to capture the first nuclear runaway 
from nuclear reactions directly.
To resolve the first runaway, the simulation requires the 
hydrodynamics timescale $t_{{\rm hyd}}$ to be smaller that the timescale 
of nuclear burning $t_{{\rm burn}}$ and convection $t_{{\rm conv}}$,
i.e. $t_{{\rm hyd}} < t_{{\rm burn}} < t_{{\rm conv}}$
\citep{Glasner2018}. Only recently there are a few pioneering models
using hydrodynamics simulations in the low-Mach number regime to follow
how the convection develops into nuclear runaway 
self-consistently \citep{Zingale2013,Jacobs2016}.
Therefore, to trace
the possible origin and to explain the observations of SN 2014J,
we considered different possibilities in how the He-detonation
can be triggered. 

On the other hand, once a detonation spot forms, the 
second detonation is in general inevitable. 
Recent three-dimensional large-scale hydrodynamics simulations
of one- or multi-spot He detonation has been found 
to be robust in triggering the off-center detonation by 
geometric convergence in a quiet He envelope \citep{Moll2013}.
The required He envelope mass required to trigger the second detonation
can drop significantly
from $\sim 0.1 ~M_{\odot}$ to $\sim 10^{-3} ~M_{\odot}$ 
for a WD mass increasing from 0.8 to 1.3 $M_{\odot}$ \citep{Fink2010},
while observationally a He envelope below 0.05 $M_{\odot}$
is favoured due to the discrepancy with the theoretical light curve
when a massive He envelope is applied \citep{Woosley2011}. 

Such diversities in the He detonation trigger and 
detonation wave interactions have provided the flexibility
to account for the diversity of SNe Ia. In \cite{Diehl2015a},
the multiple-plume structure is proposed 
to illustrate the apparent flickering of the $^{56}$Co decay line. 
Based on our models, it is possible that such feature can be 
realized by multiple spots in the He envelope. 
For a quantitative comparison, three-dimensional models 
are required for a one-one matching of the observables and predicted signatures. 
Nevertheless, using the virtue that geometric convergence and 
laminar wave shock collision do not differentiate between two- and three-dimensional
simulations, our models can shed light on what kind of shock interaction, 
and hence what kind of detonation pattern, are necessary for 
reproducing features taken from SN 2014J.

\subsection{How Typical is SN 2014J?}

In this work we explored the possible triggering 
and ignition mechanisms that might lead to
the asymmetric properties demonstrated
by the observational features of SN 2014J.
Among all models, the 
closest model we obtain is Model 110-050-R with a metallicity $\gtrsim 4~Z_{\odot}$.

The total (WD) mass required is 1.0 -- 1.1 $M_{\odot}$, at
the intermediate to high mass end of sub-Chandrasekhar mass 
WD. Stellar evolution theory suggests a progenitor mass 
constrained in a range likely to be $\sim$
6 -- 7 $M_{\odot}$ (See e.g. \cite{Catalan2008,Doherty2015} for the 
progenitor-final mass relation)\footnote{However, we remark that
at about 7 $M_{\odot}$ the final remnant mass is close to the 
transition mass of CO WD, where it is possible the core may 
have undergone advanced burning which destroys $^{12}$C 
and produced $^{20}$Ne, leaving a hybrid O+Ne+Mg core
with a C+O envelope. The exact transition mass depends on 
the stellar evolution code and input physics.}.
For a sub-Chandrasekhar mass WD to 
produce 0.6 $M_{\odot}$ $^{56}$Ni as a normal SN Ia, 
theory suggests an initial mass of $\sim 1.0 ~M_{\odot}$ 
in one-dimensional models \citep{Shigeyama1992,Nomoto2018SSR,Shen2018}. 
The $^{56}$Ni mass production is known to be sensitive to
the progenitor mass because of the density dependence of 
$^{56}$Ni production, with a minimum $> 5 \times 10^{7}$ g cm$^{-3}$.
Hence, a higher WD mass not only ensures a higher central density,
but also a higher energy release at the center, which favours the 
propagation of the detonation wave. In the one-dimensional models,
the $1.1 ~M_{\odot}$ case gives rise to a bright SN Ia
for its $\sim 0.8 ~M_{\odot}$ production of $^{56}$Ni. However, in multi-dimensional models
there are variations in the $^{56}$Ni production based on 
the initial He-detonation structure. The aspherical He-detonation
tends to give a lower $^{56}$Ni due to off-center ignition
of C+O detonation because of the density gradient as discussed in previous sections.  

The He envelope mass required by our calculations is $\sim 0.05 ~M_{\odot}$. This is 
a marginal value for the He detonation to be observed
\citep{Woosley2011}, where the optical observational features remains compatible
with the normal SN Ia data. A high He envelope mass 
likely overproduces
some iron-peak elements including Cr and V near the surface. 
This changes the typical isotope and element abundance distributions compared to normal
SNe Ia where such elements are produced in a 
deeper layer, e.g., the Chandrasekhar mass WD with 
deflagration-detonation transition. Such high opacity material in the envelope may
make the explosion appear redder in optical spectra \citep{Polin2019}. 
However, their results are based on a one-dimensional model
where the He-rich matter is always burnt from high density 
to low density. Therefore, the high density matter has
always a longer time to carry out nuclear reactions,
which favors the production of such iron-peak elements. 
However, for aspherical detonation, this is not always true.
The Cr and V production depends on how the He-detonation spread
around the He-envelope.  

Another theoretical uncertainty is the exact He mass 
when the first nuclear runaway starts. 
The exact He envelope mass depends on the mass accretion
rate and the type of binary system (single or double degenerate). 
A higher $M_{{\rm He}}$
is more likely from the double degenerate scenario while
a lower $M_{{\rm He}}$ is more likely from the single degenerate scenario. 
Such calculation has
been done by \cite{Kawai1988} from the stellar evolutionary
perspective. The 
steady state accretion of He on C+O and O+Ne+Mg WDs
are investigated in the single degenerate scenario. 
It is shown that the He envelope mass drops sharply
with the C+O core mass, with $\sim 10^{-2} ~M_{\odot}$ for a 0.7 $M_{\odot}$
C+O core down to $\sim 10^{-6} ~M_{\odot}$ for a 1.36 $M_{\odot}$
core. This shows that the steady state accretion in the
single degenerate scenario may not provide a robust way for 
accumulating a He envelope beyond $10^{-2} ~M_{\odot}$ in 
a WD of mass $1.00 ~M_{\odot}$ or above.

The detonation required at the beginning is from a ring
around the "equator". Such a configuration is shown to produce
more aspherical features in the Ni distribution, which 
would be compatible with the multiple redshifted $^{56}$Co decay lines 
measured in SN 2014J. But, we have 
also found that other types of initial detonation such as the "D"-Type
and the "B"-Type, may also produce similar characteristics, although less pronounced. 
How the He detonation is initialized is a matter of debate.

For a one-dimensional model, an entire mass shell
is ignited simultaneously because of the assumed symmetry.
However, it is unclear whether such symmetry can be 
maintained prior to the ignition. For example, in the single degenerate
scenario, the accretion of matter from the companion star
through Roche Lobe overflow has in general a high angular
momentum. Such rapidly rotating matter, when accumulated
on the stellar surface, may create strong dragging,
which disturbs the material near surface. Also, He
burning near the core-envelope interface may trigger 
convective motion \citep[see e.g.][]{Jacobs2016}. This
creates a highly turbulent background due to the shear between the
quasi-static C+O core and the rapidly rotating He-rich matter.

Knowing that the runaway of He is highly temperature-sensitive, 
it is conceivable that the ignition may occur at random locations and 
spherical symmetry is broken. In the most
extreme case, only one spot can be ignited, which corresponds to the 
"B"-Type explosion. If the rotation symmetry may be preserved, then 
the "R"-Type explosion is one of the possibilities. However, the exact configuration
will be the best estimated from the detailed multi-dimensional hydrodynamics simulations
for the last minutes before the runaway, to capture how all these 
processes interfere with each other \citep[refer e.g.][]{Zingale2011,Malone2014}. 

The early $^{56}$Ni signal can act as a tracer to the 
explosion mechanism. In Figure \ref{fig:vel_map_slice}
we show that how the He-detonation is initiated can strongly
influence the surface ejecta composition and its angular
dependence. In particular, the early low redshift $^{56}$Ni line
implies the possibility that we are observing SN 2014J close
to where shock is initialized. If we observe the shock converging point, 
the abundant elements of $^{28}$Si, $^{4}$He may easily block the 
gamma-ray. The later $^{56}$Co line can have an origin from
multiple shock convergence on the He-envelope. This can be
triggered by for example multiple rings or bubbles with non-uniform
orientation. However, the exact details may require future
study because the multiple plume feature \citep{Diehl2015a}
indicates multiple shock convergence history and different locations
for triggering the 

Our explorations favour a high metallicity of the WD compared to the 
solar metallicity. In fact, this feature is common to the observed
SNe Ia whose chemical abundance is extracted from 
their light curve and spectra. 
Such high metallicity appears to be common in recently observed 
SNe Ia, e.g. SN 2012cg \citep{Graur2016,Leung2018Chand,Leung2018SubChand}, and
SN Ia remnant 3C 397 \citep{Yamaguchi2015,Leung2018Chand}. 
These works have demonstrated that a super-solar metallicity
is paramount to boost certain isotope or element ratios, and especially the
$^{57}$Ni/$^{56}$Ni or Mn/Ni ratios. Such effects cannot be 
completely replaced
by tuning other major parameter such as the white dwarf mass, $M$, or the nuclear runaway structure.
These examples demonstrated that metallicity of exploding WDs
can be higher than solar metallicity. 
A detailed evolution path of such high metallicity WD progenitor 
would be an interesting future work. 

\subsection{Dependence on Model Dimensionality}

We note that there exist controversies regarding two-dimensional 
modeling containing symmetries, which might not necessarily be realized in reality.
Ideally, three-dimensional models are required to provide a comprehensive and
self-consistent explanation to match the explanation in a one-one correspondence. 
Here we briefly recapitulate how we use two-dimensional models 
and why this can still provide reliable estimates.

First, two-dimensional models allow more time-effective search of appropriate models.
As indicated in our previous works (Papers I and II), the parameter space which is 
suitable for SNe Ia is large. The running time for one n-dimensional hydrodynamical 
model scales with $N^{n}$ for a grid mesh of $N^n$. Typical resolution requires $\sim 500$
grids for one direction. This means three dimensional models are at minimum $\sim 500$
times more computationally expensive. It takes in general three to five days
for our two-dimensional model to complete its hydrodynamical simulation and
its nucleosynthesis. This simple scaling 
implies that a three-dimensional model requires at minimum months for a single
model. This is beyond the computational time we can afford for a practical model investigation.

Second, the large-scale aspherical effect can be well captured by two-dimensional
models. We remind that 2D-models are capable of producing three-dimensional 
aspherical structure such as a bubble or a ring naturally, and also the one-dimensional
spherical structure. Three-dimensional simulations can produce more complex 
structures in the form of multiple bubbles, for example. 
In fact, the processes determining how the second detonation starts 
depends on the wave collision details. These can already be captured by 
one of the two-dimensional scenarios. Furthermore, two-dimensional models
provide the minimum perturbation from the spherical symmetry. Notice that 
SN 2014J has features which can be explained by the classical W7 model
on a broader picture, even though having aspherical features.
meanwhile. Therefore, it might be considered as a generalization 
to start from models which behave almost spherical, and extend them to lower symmetries. 

Third, the symmetry is conserved in the simulation. As demonstrated in \cite{Moll2013}, where
one-, two- and three-dimensional simulations of the sub-Chandrasekhar mass models are carried
out, one of their explicit three dimensional models with a two-dimensional counterpart gives
agreeing results with each other.
This provides some support that in the explosion phase, symmetry does 
not break during its propagation. Similar three-dimensional models in \cite{Gronow2020}
also demonstrated similar features, in that the detonation propagates like a two-dimensional
front. Furthermore, in this work, we further show that the 
two-dimensional spherical models explode spherically as in the one-dimensional case.
Both results support that our axis-symmetric model remains to be axis-symmetric throughout
the simulations, as long as turbulent motion is unimportant. This is true during the 
explosion phase ($\sim$ 1 s). When time is sufficiently long, i.e. in the nebular phase, 
we expect that the Rayleigh-Taylor instabilities play a role and perturb the morphology.
Then, initial seeds break the rotation symmetry. 
However, such effects are secondary compared to the large-scale asymmetry and 
require more time to grow.
Also, similar to models in the literature \citep[see e.g.][]{Fink2007,Sim2010,Fink2010},
we set up WDs in hydrostatic equilibrium as the initial condition. The quiet environment suggests
that the turbulent motion is suppressed.

Finally, in this work we focus on the common features which exist in both
two- and three-dimensional models. In particular we investigate how shock waves superpose, interact
with each other, or grow by themselves through geometric convergence, and consequently generate the 
structure that breaks the spherical symmetry. As indicated in Paper II, 
how the wave interacts are independent of the boundary condition.
Thus, our two-dimensional model can offer the necessary starting point
to explore which kind of detonation structure is necessary to generate
the corresponding interaction for creating the large-scale asymmetry.

\subsection{Conclusion}

In this article we explored the parameter space in the classical double detonation model which can
produce observables indicating deviations from spherical symmetry.
We studied how the initial detonation geometry affects the final explosion morphology by 
examining the ejecta composition in the spatial distribution and velocity space. 
We studied how the spherical symmetry can be broken for creating large-scale asymmetry. 
The sub-Chandrasekhar mass WD progenitor tends to produce more pronounced asymmetry than the 
Chandrasekhar mass WD progenitor. The surface He detonation can be the origin of the early $^{56}$Ni gamma-ray line of some SNe Ia, e.g. SN 2014J and the recently observed early bumps in the 
observed light curves of some SNe Ia.

We have examined how the initial mass, He-detonation geometry, affects the final explosion results, 
in particular the ejecta geometry and element distribution in both spatial and velocity
phase space. We observe that starting the He-detonation as a bubble (with lowest symmetry), 
to a ring, and then a sphere (with highest symmetry), may give observable differences 
in the ejecta morphology, ejecta velocity for the characteristics elements, 
including He, O, Si, S, Fe, and Ni, and their directional dependence. 

We have provided a detailed case study on searching for models which may resemble with 
the qualitative features observed in SN 2014J based on the gamma-ray line detections 
and the late-time photometry of the optical band. Four key aspects of SNe Ia explosion are:

\noindent (1) The total mass $M$ of the WD determines the total 
$^{56}$Ni production; 

\noindent (2) The He-envelope mass $M_{{\rm He}}$ determines the 
large scale asymmetry in the radial distribution 
of $^{56}$Ni;

\noindent (3) The metallicity determines the required $^{57}$Ni/$^{56}$Ni 
mass;

\noindent (4) The initial He-runaway geometry determines the 
surface $^{56}$Ni distribution.

From our explorations simulating a set of key scenarios, 
we conclude that the SN 2014J progenitor should have the following properties:

\noindent (1) An initial He-detonation in the orbital plane set by the binary companion;
 
\noindent (2) a WD mass in the range from 1.00 to 1.10 $M_{\odot}$;

\noindent (3) a WD metallicity in the range from $3$ to $5 ~Z_{\odot}$;

\noindent (4) a He envelope mass $\sim 0.05$ -- $0.10~M_{\odot}$.

We also derived the detailed velocity distributions of some major isotopes, 
for example $^{16}$O, $^{28}$Si, $^{54}$Fe and $^{56-58}$Ni and the spatial distributions
of major IPEs including Cr, Mn, Fe and Ni. Future observations of the ejecta morphology by 
specific elements (e.g. \cite{Seitenzahl2019}) can provide a strong constraint on the models 
presented in this work. Large-scale features in these objects might reveal how the detonation 
has interacted during its propagation, thus shedding light on its initial detonation pattern.

At last we discuss the recent application of SN tomography as presented in \cite{Seitenzahl2019} for the 
SN remnants 0519-69.0, 0509-67.5 and N103B. From how aspherical the SN ejecta in the reverse-shock 
heated region, and their corresponding Fe-mass are, we can deduce the fundamental properties of the
progenitor including whether it is a Chandrasekhar or sub-Chandrasekhar WD, the expected initial
mass and detonation geometry. We also summarize the $M$- and $Z$-dependence of the major element 
ratios typically found in spectra of SN remnants.  

\acknowledgments
This work has been supported by the World Premier International
Research Center Initiative (WPI Initiative), MEXT, Japan.
S.C.L. acknowledges support from grant HST-AR-15021.001-A and 80NSSC18K1017.
K.N. acknowledges support from JSPS
KAKENHI Grant Numbers JP17K05382 and JP20K04024.
Thomas Siegert is supported by the German Research Society (DFG-Forschungsstipendium SI 2502/1-1).
This research was also supported by the DFG cluster of excellence "Origin and Structure 
of the Universe".

\appendix

\section{White Dwarf Models for Type Ia Supernova}

Type Ia supernovae (SNe Ia) are the thermonuclear explosions of
 CO white dwarfs (WDs) \citep[see e.g.][]{Hillebrandt2000,Nomoto2017}. 
A single CO WD does not spontaneously undergo nuclear burning.
In a close binary system, on the other hand, the WD gains mass by mass
transfer from its companion star, which includes a slightly evolved near main-sequence star, a red-giant,
and a He-star \citep[single degenerate scenario, e.g.,][]{Nomoto1982a,Kawai1988}
or a WD \citep[double degenerate scenario][]{Iben1984,Webbink1984}.

During the mass accretion, if the accretion rate is relatively low,
the accreted He is accumulated on the surface and eventually He-burning is
ignited first in the off-center hot spot when the WD mass is the
sub-Chandrasekhar mass \citep{Nomoto1982a}.  Then He-detonation is
developed \citep{Nomoto1982b}.

If the accretion rate is relatively high, on the other hand, the WD
mass reaches the near-Chandrasekhar mass and the central temperature
exceeds $\sim 3 \times 10^8$ K.  Then the energy generation rate of
$^{12}$C-burning exceeds the neutrino cooling rate.  $^{12}$C-burning
is unstable to develop a C-deflagration in the center
\citep{Nomoto1982a}
\footnote{
Depending on the mass accretion rate, the white dwarf can undergo
centered deflagration, double detonation or accretion-induced collapse
\citep[see also][for a detailed phase diagram]{Nomoto1982b,Nomoto1984}}.

\subsection{Near-Chandrasekhar Mass White Dwarf Models}

When $^{12}$C-burning is ignited in the center of the
near-Chandrasekhar mass WD, the central density is as high as $2 - 3
\times 10^9$ g cm$^{-3}$.  Electrons are strongly degenerate, so that
the gas pressure is not so sensitive to the temperature.  The
temperature rise becomes unregulated as the carbon burning rate is
strongly temperature sensitive ($\sim T^{33}$).  This sets the first
seed of nuclear runaway.  Simultaneously, the rapid temperature rise
does not trigger a shock because the pressure growth is small compared
to the temperature rise due to the strong degeneracy even when the
central temperature becomes as high as $\sim 10^{10}$ K
\citep{Nomoto1976,Nomoto1977}.  As a result, the temperature jump
becomes a localized event, where the temperature discontinuity
propagates by microscopic processes instead of macroscopic shock
compression.  Thus, the initial runaway is likely to be in the form of
a deflagration wave \citep{Nomoto1976,Nomoto1984}.  The short mean
free path of electron conduction in such a density implies very thin
flame front ($\sim 10^{-3}$ cm) compared to the size of a WD ($\sim
10^3$ km) \citep{Timmes1992a}.

Despite the turbulent motion emerges down to the Kolmogorov scale
($10^{-3}$ cm assuming a typical Reynolds number of $\sim 10^{14}$),
the Gibson scale decreases with density and it is in general larger
than the Kolmogorov scale.  Flame structure with a size below the
Gibson scale ($\sim 10^{-1}$ km at the center), is smoothed
\citep{Niemeyer1995a,Roepke2003,Roepke2004a}, albeit eddies can appear
below this scale.

If the propagation speed of the subsonic deflagration is fast enough,
the deflagration efficiently releases the necessary energy for
unbinding a WD and creates a successful explosion like W7
model\citep{Nomoto1984} as also does the detonation wave
\citep{Arnett1969}.

The explosive nuclear burning at high densities synthesize iron-peak elements
\citep[e.g.,][]{Thielemann1986,Iwamoto1999}. However, the observed intermediate mass
elements should be synthesized at lower densities, which suggests the
explosion consists of a subsonic burning, i.e. deflagration
\citep{Nomoto1976,Nomoto1977}, which decreases the densities at the
flame front.

If the propagation of the deflagration wave is slow, it may not unbind
the star (but in some cases may cause pulsation)
because the stellar expansion makes the deflagration wave quench \citep{Nomoto1976}.
Subsequent transition from deflagration to detonation is vital
for explaining a successful SN Ia explosion, known as the 
deflagration-detonation transition
\citep{Nomoto1984,Khokhlov1991a,Arnett1994a,Arnett1994b,Iwamoto1999,Gamezo2003,Gamezo2004,Roepke2007a}.
However, the deflagration-detonation transition requires a turbulence strength which is 
less likely to be reached \citep[see, e.g.,][]{Khokhlov1997,Niemeyer1999,Lisewski2000,Gamezo2005,Woosley2009}.
In order to realize this effect, ab initio numerical 
experiments with very fine resolutions ($\sim 0.1$ km) are necessary \citep{Kushnir2012},
which are one to two orders of magnitude below the affordable resolution.
Recent direct experiments, both numerical and laboratory
ones using methane-air mixture resolving turbulent motion explicitly, 
have demonstrated that turbulent acceleration
can be an important key factor \citep{Poludenko2011,Poludenko2019}. 

The motion of the deflagration wave can be convoluted.
The subsonic propagation ($\sim 1 \%$ speed of sound) implies that 
the deflagration wave structure is coupled with the underlying
fluid motion, which means that the flame structure is also susceptible to 
various hydrodynamics instabilities such as 
the Rayleigh-Taylor instabilities \citep{Bell2004b,Zingale2005,Hicks2015,Hicks2019},
Kelvin-Helmholtz instabilities, 
Landau-Derrieus instabilities \citep{Bell2004a,Roepke2004b}
and pulsational instabilities \citep{Glazyrin2013a, Glazyrin2013b, Glazyrin2014, Poludenko2015}.
On the contrary, the supersonic detonation is less sensitive to 
fluid motion. However, direct numerical simulations of the small scale
detonation shows spontaneous cellular structure formation behind the
detonation wave front \citep{Gamezo1999}.

To model the explosion, following how the deflagration propagates
reveals how the energy is released. The sub-grid scale of the reaction front
indicates that on-site modeling is inaccurate, but a sub-grid scale model
is necessary to describe partial cell burning and irregular wave front
inside the cell. This relies on the sub-grid scale turbulence model
\citep{Clement1993,Shih1995b,Shih1995a,Niemeyer1995a} and 
the flame tracking scheme.

The sub-grid scale turbulence model assumes that eddy motion below the resolved scale
can be well described by statistical models. This gives an accurate approximation
given the large difference between the resolved scale and the much smaller
Kolmogorov's scale. Scaling relation has been studied explicitly in direct 
simulation \citep[see e.g.][]{Fisher2019}.
The model tracks the generation and dissipation of 
eddy motions by channels including shear-stress, compression,
Rayleigh-Taylor instabilities and so on 
\citep[see e.g.][]{Shih1995b,Shih1995a,Niemeyer1995a,Schmidt2006b}.

Flame tracking schemes are algorithms designed for resolving 
sub-grid scale features. There are multiple representations, including 
(1) the advective-diffusive-reactive
equations \citep{Khokhlov1995,Vladimirova2006,Townsley2007}, 
(2) level-set methods \citep{Osher1988,Sethian2001}, and
(3) point-set methods \citep{Glimm1999,Glimm2000, Zhang2009, Leung2015a}.
The main idea is to introduce additional variables with model parameters which 
represent how much the grid is partially burnt, from that the actual
flame front geometry is reconstructed.

\subsection{Sub-Chandrasekhar Mass White Dwarf Models}

In a sub-Chandrasekhar mass WD, the less degenerate matter with a lower density
means that the detonation is more likely \citep{Nomoto1982a}. Such scenario is viable when 
the surface energy production is faster than its heat loss 
by convection or expansion \citep{Jacobs2016}.
The initial nuclear runaway can be triggered by accretion from its companion star
in the single degenerate scenario, or through a violent merger
in the double degenerate scenario 
\cite[see, e.g.][]{Tanikawa2015,Tanikawa2019}.
Pure CO matter has a high ignition threshold for its high temperature
($\sim 1 - 2 \times 10^9$ K) \citep[e.g.,][]{Sato2015}, which is shown to be 
difficult to trigger and sensitive to the way of contact \citep{Dan2012}.
WD merger with a helium envelope can suppress this ignition condition,
but its required hotspot size can be non-realizable in a thin white dwarf
envelope \cite[see e.g.][]{Shen2009}. Mixing with C/O-rich matter through for example turbulence can resolve this difficulty \citep{Holcomb2013,Piro2015}.
The violent merger of two CO white dwarfs \citep{Pakmor2012}
are therefore challenging for a robust ignition 
as the collision can fail to generate the 
spots, sufficiently hot for the first runaway \citep{Dan2012,Dan2014}
Mixture with helium provided by its He-envelope or the companion star lowers the 
ignition temperature such that the detonation trigger is less sensitive
to the merging dynamics \citep{Shen2014}. A thin layer of He ($\sim 0.01 ~M_{\odot}$) can already
trigger the second explosion more robustly \citep{Pakmor2013}.
When the He-detonation fails to trigger the second detonation. 
The star develops like a nova and explodes as a so-called Type ".Ia" supernova 
\citep{Bildsten2007,Shen2010,Waldman2011}.

Even when the C-detonation trigger becomes robust with the aid of He, 
the exact position and timing of the detonation are unclear because they depend
on the dynamics of the He-atmosphere. These require multi-dimensional 
low Mach number simulations of the atmosphere for multiple eddy turnover time. 
Multiple possibilities exist including direct He-ignition, C-ignition, 
or ignition after the merger process when a Chandrasekhar mass WD 
is formed \citep{Dan2011,Shen2018,Tanikawa2019}. 
Geometric convergence in a low-mass WD is more difficult
to achieve \citep{Shen2014}. High resolution simulations using ab initio approach
is necessary to trace when and where the first hot spot appears \citep{Fenn2017}. 
The asymmetry in a three-dimensional simulation tends to suppress 
the prompt detonation as the geometrical convergence breaks down in 
the violent merger scenario \citep{Fenn2016}.

In a low mass WD, the detonation front has a size comparable with the resolved scale.
This allows directly coupling of the hydrodynamics with a nuclear reaction
network \citep{Shen2018,Polin2019}.
In a more massive CO WD (central density $\geqslant 10^8$ g cm$^{-3}$),
the detonation width can be much smaller than the resolved
grid size ($\sim 10$ km). sub-grid scale methods or adaptive mesh refinement
are often used in the literature.

\bibliographystyle{aasjournal}
\pagestyle{plain}
\bibliography{biblio}

\end{document}